\documentclass[journal=jpclcd,manuscript=article]{achemso}
\setkeys{acs}{email=false}
\usepackage[english]{babel}
\usepackage[T1]{fontenc}
\usepackage{enumerate}
\usepackage{csquotes}
\usepackage[fleqn]{amsmath}
\usepackage{amsmath}%
\usepackage{MnSymbol}%
\usepackage{wasysym}%
\usepackage{soul}
\usepackage{relsize}
\usepackage{setspace}
\usepackage{titlesec}
\usepackage{graphicx}
\usepackage{subcaption}
\usepackage{pdfpages}
\graphicspath{{./figures/}}

\title{Quantifying Confidence in DFT Predicted Surface Pourbaix Diagrams and Associated Reaction Pathways for Chlorine Evolution}

\author{Vaidish Sumaria}
\affiliation{Department of Chemical Engineering, Carnegie Mellon University,
Pittsburgh, PA 15213, USA}
\alsoaffiliation{Department of Mechanical Engineering, Carnegie Mellon University,
Pittsburgh, PA 15213, USA}

\author{Dilip Krishnamurthy}
\affiliation{Department of Mechanical Engineering, Carnegie Mellon University,
Pittsburgh, PA 15213, USA}

\author{Venkatasubramanian Viswanathan}
\affiliation{Department of Mechanical Engineering, Carnegie Mellon University,
Pittsburgh, PA 15213, USA}
\alsoaffiliation{Department of Chemical Engineering, Carnegie Mellon University,
Pittsburgh, PA 15213, USA}
\email{venkvis@cmu.edu}
\begin{document}
\singlespacing
\begin{tocentry}
\includegraphics[width=2in]{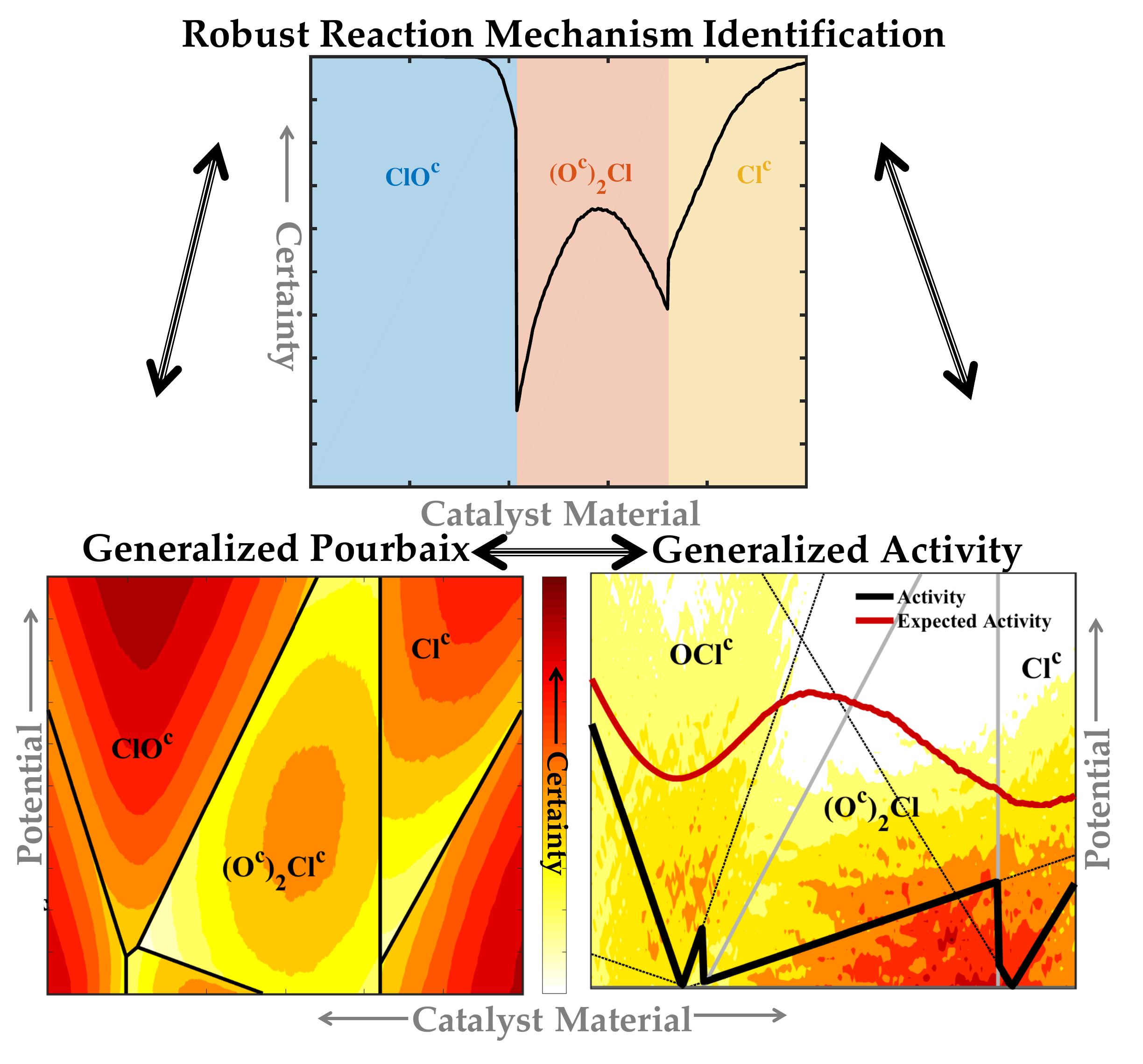}
\end{tocentry}

\begin{abstract} 
Catalytic activity predictions and the identification of active sites rely on precisely determining the dominant reaction mechanism. The activity governing mechanism and products could vary with the catalyst material, which can be described by material descriptor(s), typically the binding strength(s) of key intermediate species. Density functional theory calculations can be used to identify dominant reaction mechanisms. However, the dominant reaction mechanism is sensitive to choice of the exchange correlation functional.  Here, we demonstrate using the example case of chlorine evolution reaction on rutile oxides, which can occur through at least three reaction mechanisms each mediated by different surface intermediates and active sites. We utilize Bayesian error estimation capabilities within the BEEF-vdW exchange correlation (XC) functional to quantify the uncertainty associated with predictions of the operative reaction mechanism by systematically propagating the uncertainty originating from DFT-computed adsorption free energies. We construct surface Pourbaix diagrams based on the calculated adsorption free energies for rutile oxides of Ru, Ir, Ti, Pt, V, Sn and Rh. We utilize confidence-value (c-value) to determine the degree of confidence in the predicted surface phase diagrams. Using the scaling relations between the adsorption energies of intermediates we construct a generalized Pourbaix diagram showing the stable surface composition as a function of potential and the oxygen binding energy on the cus site ($\Delta E_{O^c}$). This is used to consistently determine activity volcano relationships. We incorporate the uncertainty in linear scaling relations to quantify the confidence in generalized Pourbaix diagram and the associated activity. This allows us to compute the expectation limiting potential as a function of $\Delta E_{O^c}$, which provides a more appropriate activity measure incorporating DFT uncertainty. We show that the confidence in the classification problem of identifying the active reaction mechanism is much better than the prediction problem of determining catalytic activity.  We believe that such a systematic approach is needed for accurate determination of activities and reaction pathways for multi-electron electrochemical reactions such as N$_2$ and CO$_2$ reduction.
\end{abstract}

\section{Introduction}
A major challenge in surface electrocatalysis involves identifying the dominant reaction mechanism from the many possible competing electrochemical reactions at the electrode/electrolyte interface.~\cite{BockrisSurfaceElectrochemistry} This is further complicated by the dynamic state of the surface depending on the external conditions, for e.g. electrode potential, pH, aqueous electrolyte.\cite{hansen2008surface,yan2017surface,bajdich2013theoretical,hansen2010electrochemical} The coupling between surface state and dominant reaction mechanism is crucial for most important electrochemical reactions, for e.g. chlorine evolution,~\cite{hansen2010electrochemical} hydrogen evolution,~\cite{skulason2007density} nitrogen reduction,~\cite{montoya2015challenge} and carbon-dioxide reduction.~\cite{kuhl2012new, hori1989formation,hori1997electrochemical,schouten2011new,dewulf1989electrochemical,peterson2010copper, schouten2012two}

Experimentally, macrokinetic measurements, to determine reaction orders, pH dependence and Tafel slope, can provide useful insights but typically several competing mechanistic pathways could still provide a plausible explanation for all the data.\cite{krishtalik1981kinetics,hansen2010electrochemical} An emerging frontier is using spectroscopic methods to identify stable surface species, for e.g., infrared spectroscopy, ambient pressure XPS in conjunction with electrochemical measurements to identify the dominant mechanistic pathways.~\cite{wieckowski2011contrast, mukerjee1995role, 1996i, dewulf1989electrochemical,wakisaka2009identification} 

First-principles calculations have been used to identify probable reaction mechanisms through a combination of surface Pourbaix diagrams and free energy diagrams for the different mechanistic pathways.~\cite{su2012identifying,hansen2010electrochemical}  However, it is well-known that DFT calculations have finite accuracy; hence, a key question emerges on the likelihood of the identified dominant reaction pathways.  Systematic progress has been made possible with the incorporation of Bayesian error estimation capabilities\cite{mortensen2005bayesian} into exchange correlation functionals. This has been used to quantify uncertainty associated with reaction rates in heterogeneous catalysis \cite{medford2014assessing} and electrocatalysis.~\cite{deshpande2016quantifying}  More recently, we showed that uncertainty quantification can be used as a tool to decide the most appropriate descriptor for computational predictions.\cite{krishnamurthy2018maximal}  However, these analyses make simplistic assumptions about the state of the surface and do not incorporate a self-consistent loop necessary to determine surface reactivity.

In this work, we demonstrate an approach to accurately incorporate the self-consistent loop between surface state and reaction pathways while systematically incorporating uncertainty associated with chemisorption energies. We apply this approach to an important electrochemical reaction, chlorine evolution.  This reaction is chosen for its industrial importance\cite{OBrien2005} and the complexities involved in mapping the selectivity with respect to oxygen evolution reaction.~\cite{trasatti1984electrocatalysis} Prior work by Hansen et al. \cite{hansen2010electrochemical} shows that the overpotential for oxygen evolution to occur on rutile oxides, which involves three oxygen intermediates, is always higher than that for chlorine evolution.~\cite{man2011universality}  This can be rationalized based on the fact that finding a catalyst material for an electrochemical reaction involving more elementary steps is generally challenging owing to inherent scaling relations.~\cite{viswanathan2014unifying} In this work, we focus on the reaction mechanisms for the chlorine evolution reaction on rutile oxides. In this analysis we begin with constructing the Pourbaix diagram for RuO$_2$, IrO$_2$, TiO$_2$, PtO$_2$, VO$_2$, SnO$_2$ and RhO$_2$.  Inspired by the work of Hansen et al.,~\cite{hansen2010electrochemical} we determine a generalized Pourbaix diagram with the activity volcano relationships of the three reaction mechanisms to develop a generalized trimodal activity relationship. We use the previously defined quantity, c-value, to quantify confidence in surface phase diagrams. Through the error-estimation approach, an ensemble of generalized Pourbaix diagrams are constructed that lead to an ensemble of generalized activity relationships. We compute the expected activity for a range of material descriptor values, which is the probability weighted average of the predicted limiting potentials. Finally, we utilize this determine the likelihood of reaction mechanisms for different material descriptor values. We quantify this through a variant of the c-value metric to determine the reaction mechanism prediction confidence.   An important conclusion is that confidence associated with reaction mechanisms is much greater than the predicted activity values.  We believe this will be crucial to accurately understand different reaction mechanisms and thereby the activity and the nature of the active site.

\section{Methods}
\subsection{Calculation details}
Calculations were performed using the projector augmented-wave (PAW) method\cite{mortensen2005real} as implemented in the GPAW package\cite{enkovaara2010electronic} using the Bayesian error estimation functional with van der Waals correlation (BEEF-vdW) which has built-in error estimation capabilities.\cite{wellendorff2012density}  The functional utilizes an ensemble of exchange correlation (XC) functionals resulting in an ensemble of energies from which the uncertainty in the adsorption energies can be calculated.\cite{medford2014assessing,deshpande2016quantifying} A periodically repeated 4-layer slab is chosen for all the considered rutile oxide (110) surfaces of RuO$_2$, IrO$_2$, TiO$_2$, PtO$_2$, RhO$_2$, SnO$_2$ and VO$_2$. A 4$\times$4$\times$1 Monkhorst-Pack type k-point grid was used for a 2$\times$1 surface unit cell with two bridge and two cus ($^b$ and $^c$ respectively) sites. The two bottom layers of the unit cell were kept fixed and the top two layers with the adsorbates were allowed to relax with a force criterion of < 0.05 eV/{\AA}. A Fermi smearing of 0.01 eV is used and all calculated energies are extrapolated to an electronic temperature of 0 K.

Following the work of Hansen et al. \cite{hansen2010electrochemical}, we consider adsorption of the intermediates OH$^b$ and O$^b$ at the bridge site, and adsorption of O$^c$, OH$^c$, Cl$^c$, OOH$^c$ and OCl$^c$ at the coordinatively unsaturated site (cus), and the adsorption of O$_2^{cc}$ and (O$^c$)$_2$Cl at adjacent cus sites. All relevant combinations involving the two adsorption sites and the various adsorbates were considered. For the range of reaction conditions relevant to chlorine evolution, the bridge sites bind intermediates relatively strongly and are thereby covered with oxygen. The adsorption energies of the considered adsorbates are referenced to chlorine and hydrogen gas. \cite{hansen2010electrochemical} We use gas phase H$_2$O as the reference state for oxygen, by assuming chemical equilibrium with liquid water at 298 K and 0.035 bar \cite{norskov2005trends}, to avoid the well-known errors made by DFT in describing O$_2$. We assume a negligible effect on the energetics due to the interaction between the electrolyte constituents and the surface since it is known that rutile oxides are water repelling, which has been confirmed by small (<0.05 eV) changes in O and OH adsorption energies on incorporation of water layer structures on RuO$_2$ \cite{rossmeisl2007electrolysis}. Additionally, the effect of electric field in the Helmholtz layer is not taken into account since its effect is negligible for adsorbates with small dipole moments perpendicular to the surface.\cite{karlberg2007estimations}

\subsection{Quantifying confidence in predicted surface states} \label{ss:cval_pourbaix}
Pourbaix diagrams represent the landscape of the most thermodynamically stable state (minimum Gibbs free energy of adsorption) of a given surface over a range of operating potentials (U) and pH values. The ensemble of functionals results in an ensemble of Pourbaix diagrams, allowing us to obtain a measure of the confidence in a predicted surface state by quantifying the agreement between functionals. More precisely, we use the confidence-value\cite{houchins2017quantifying} (c-value), which in this context can be defined as the fraction of the ensemble that is in agreement with the hypothesis of the best-fit (or optimal BEEF-vdW) functional, and is given by
\begin{equation}
c(U,pH) =  \frac{1}{N_{ens}} \sum_{n=1}^{N_{ens}} \prod_{s_i\neq s_{opt}} \Theta(\Delta G_{s_i}^n(U,pH)-\Delta G_{s_{opt}}^n (U,pH))
\end{equation}
\noindent where, $s_i \in S$, the set of all considered surface states, and $s_{opt}$ is the thermodynamically stable surface state predicted by the BEEF-vdW optimal functional at a given $U$ and $pH$. $\Delta G^n_{s_i}$ refers to the adsorption free energy of the $i^{\mathrm{th}}$ surface state given by the $n^{\mathrm{th}}$ member of the ensemble of functionals. $\Theta(x)$ denotes the Heaviside step function.

\subsection{Reaction mechanisms and expected activity} \label{ss:UEL}
We define the activity in terms of the limiting potential ($U_L$), given by the lowest potential at which all the involved reaction steps are downhill in free energy. The thermodynamic analysis forms a necessary criterion but could be insufficient as we do not incorporate activation energies in the study due to the associated computational challenges in calculating them. It has been shown that this analysis remains
consistent with a wide range of experiments on metals and
alloys for oxygen reduction even when kinetics is taken into account.~\cite{hansen2014unifying}  Further, this framework has been used to rationalize trends in reactivity for oxygen reduction,~\cite{viswanathan2012universality} oxygen evolution,~\cite{man2011universality} and hydrogen evolution.~\cite{greeley2006hydrogen}

The activity of rutile oxides is a function of the reaction pathway by which chlorine evolution could occur. We consider all the possible reaction mechanisms based on the thermodynamically stable chlorine containing reaction intermediates:
\begin{enumerate}[I.]

\item \textbf{Pathway mediated by the intermediate $\mathbf{ClO^c}$} \label{rxn1}

$O^c \ + \ Cl^- (aq.) \ \rightarrow \ ClO^c \ + \ e^- $ \\
$ClO^c \ + \ e^- \ + \ Cl^- (aq.) \ \rightarrow \ O^c \ + \ Cl_2 (g) \ + \ 2e^-$

The limiting potential for this mechanism can be given as:

$U_L \ = \ U^{eq} \ + \ |\Delta G(ClO^c) - \Delta G(O^c)|/e $

\item \textbf{Pathway mediated by the intermediate $\mathbf{Cl(O^c)_2}$} \label{rxn2}

$O_2^{cc} \ + \ Cl^- (aq.) \ \rightarrow \ Cl(O^c)_2 \ + \ e^- $ \\
$Cl(O^c)_2 \ + \ e^- \ + \ Cl^- (aq.) \ \rightarrow \ O_2^{cc} \ + \ Cl_2 (g) \ + \ 2e^-$

The limiting potential for this mechanism can be given as:

$U_L \ = \ U^{eq} \ + \ |\Delta G(Cl(O^c)_2) - \Delta G(O_2^{cc})|/e $

\item \textbf{Pathway mediated by the intermediate $\mathbf{Cl^c}$}  \label{rxn3}

$ Cl^- (aq.) + ^c \ \rightarrow \ Cl^c \ + \ e^- $ \\
$Cl^c \ + \ e^- \ + \ Cl^- (aq.) \ \rightarrow \ ^c \ + \ Cl_2 (g) \ + \ 2e^-$

The limiting potential for this mechanism can be given as:

$U_L \ = \ U^{eq} \ + \ |\Delta G(Cl^c)|/e$
\end{enumerate}

We compute the expected limiting potential, $U_{EL}(\Delta E_{O^c)}$, which is determined as the probability-weighted average of the limiting potential distribution, given by 
\begin{equation}
U_{EL} =  E[U_L] = \int_{U^{min}_L}^{U_{L}^{max}} U_L\ p(U_L)~dU_L
\end{equation}
This approach relies on computing the probability distribution (Figure \ref{fig:UEL}) of the predicted limiting potentials for chlorine evolution using the ensemble of the predicted activity volcanoes and the associated generalized phase diagrams (refer to section \ref{LPEAP}). 

\subsection{Quantifying confidence in the predicted reaction mechanism} \label{ss:cval_rxn_mech}
The predicted active reaction mechanism for chlorine evolution varies with materials as a function of the chosen material descriptor, represented as $m_{pred}(\Delta E_{O^c})$, which maps any given value of $\Delta E_{O^c}$ to the corresponding reaction mechanism prediction from the set of possible mechanisms denoted by \{0,1,2,...,i,...,n\}, where $i$ denotes the $i^{\mathrm{th}}$ mechanism and $i=0$ indicates no active mechanism. 
For chlorine evolution on rutile oxides, $n=3$ as there are three distinct reaction mechanisms involved (refer to section \ref{ss:UEL}).  For materials where the reaction mechanism is mediated through $\mathrm{ClO^c}$, we assign $m_{pred} = 1$.  Similarly, when the mechanism is mediated through $\mathrm{Cl(O^c)_2}$ and $\mathrm{Cl^c}$, we assign $m_{pred} = 2$ and $m_{pred} = 3$, respectively.

We quantify the confidence in the predicted reaction mechanism as a function of $\Delta E_{O^c}$ through a Bayesian error-estimation approach similar to that outlined in section \ref{ss:cval_pourbaix} for computing the confidence in predicted surface states. At each value of $\Delta E_{O^c}$, the mechanism prediction confidence $c_{m_{pred}}(\Delta E_{O^c})$ is calculated as the fraction of the ensemble of functionals that is consistent with the predicted active mechanism based on the best-fit functional, $m_{pred}^{opt}(\Delta E_{O^c})$.  
\begin{equation}
c_{m_{pred}}(\Delta E_{O^c}) =  \frac{1}{N_{ens}} \sum_{n=1}^{N_{ens}} \delta (m_{pred}^n(\Delta E_{O^c})-m_{pred}^{opt}(\Delta E_{O^c}))
\end{equation}
\noindent where, $n$ denotes the $n^{th}$ functional, $N_{ens}$ is total number of functionals in the ensemble and $\delta(x)$ denotes the Dirac delta function. 

In regimes where the $c_{m_{pred}}$ value is lower than 1, it becomes important to determine whether the reaction mechanism predicted by the majority of functionals agree with the optimal functional. This can be extended to determine a measure of confidence in any given reaction mechanism or the $i^{\mathrm{th}}$ reaction mechanism $m_i$ being operative.  For example, for the mechanism mediated by $\mathrm{ClO^c}$, $c_{m_{pred}=1}(\Delta E_{O^c})$, determines the fraction of functionals that predict that this mechanism is active.  A generalized relation can be given as,
\begin{equation}
\label{cmechi}
c_{m_{pred}=i}(\Delta E_{O^c}) = \frac{1}{N_{ens}} \sum_{n=1}^{N_{ens}}\delta (m_{pred}^n(\Delta E_{O^c})-i).
\end{equation}

\section{Results and Discussion}
\subsection{Confidence in stable surface phase predictions}
\begin{figure}[t!]
	\includegraphics[width=\textwidth]{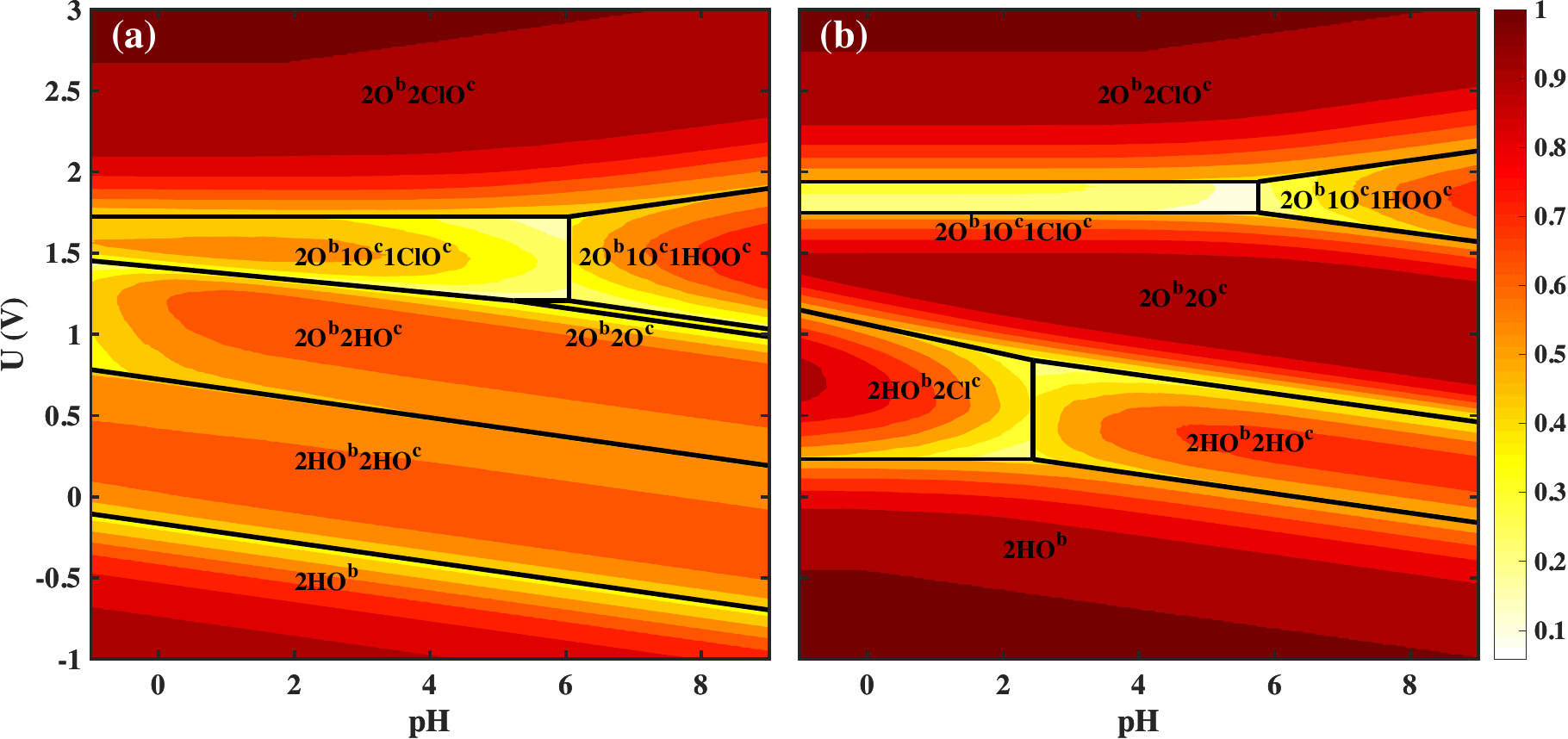}
    \caption{Surface Pourbaix diagrams for (a) IrO$_2$ and (b) RuO$_2$. The bridge and cus sites of adsorption on rutile (110) oxides have been denoted by $\mathrm{^b}$ and $\mathrm{^c}$ respectively. The black lines define the surface phase boundaries and the stable surface states are depicted using the surface intermediate species. The two Pourbaix diagrams have been constructed for catalyst surface in equilibrium with Cl$^-$, H$^+$ and H$_2$O at 298.15 K and a$\mathrm{_{Cl^+}}$=1. For both IrO$_2$ and RuO$_2$, we observe that chlorine evolution is mediated using the reaction mechanism involving the stable intermediate ClO$\mathrm{^c}$. The uncertainty in the surface phase predictions have been quantified using the c-value, which we define as a metric for prediction confidence in the context of Pourbaix diagrams. The c-values of the associated predictions have been shown using a color-map. Phase boundaries are characteristic of regions with low c-values, identifying electrochemical operating regimes where DFT predictions are subject to higher uncertainty. Surface Pourbaix diagrams with associated c-values for PtO$_2$, TiO$_2$, RhO$_2$, SnO$_2$ are included in  Figure S3 of the supporting information.}
    \label{fig:UvspH}
\end{figure}

In this section, we attempt to answer the important question of the level of confidence in the predicted stable surfaces states for the active rutile oxides, using the approach outlined in section \ref{ss:cval_pourbaix}. Figure \ref{fig:UvspH}(a) depicts the surface phase diagram for IrO$_2$ with the associated c-values indicating the likelihood of occurrence of the predicted surface Pourbaix diagram. Although there exist some differences in the positions of phase boundaries, we observe that the predicted set stable surface states are consistent with that reported by Hansen et al.~\cite{hansen2010electrochemical} The quantification of confidence in the predicted Pourbaix diagram allows us to identify regimes of high confidence as well as those with high uncertainty. We notice that regions predicted with low c-values are close to surfaces phase boundaries. This identifies electrochemical operating regimes over which activity predictions based on the associated stable phase are subject to high uncertainty and require higher fidelity computations in conjunction with experimental validation. 

A highly active catalyst for chlorine evolution entails a $\Delta G \sim 0$ eV near $\mathrm{U=1.36\ V}$ for the formation of Cl$^c$. On IrO$_2$, we find that an active mechanism involving Cl adsorbed directly on an Ir cation is not operative for chlorine evolution. However, we observe that for a large range of $pH$ ($0 \lesssim pH\lesssim 6$), ClO$^c$ on the surface is thermodynamically stable for U>1.5 V. This suggests that chlorine evolution on IrO$_2$ could be mediated by reaction pathway \ref{rxn1}, as described in section \ref{ss:UEL}, consistent with the predictions using the RPBE XC.\cite{hansen2010electrochemical}

On RuO$_2$ (Figure \ref{fig:UvspH}(b)), we find that the predicted stable surface phases in this work compares well with that reported by Hansen et al. at low potentials ($U\lesssim 1$ V).~\cite{hansen2010electrochemical} In this range of potentials, we observe a transition from only bridge sites being covered with OH to being completely covered by OH with increasing U. At higher potentials ($\sim$ $1\lesssim U\lesssim 1.8$ V), we find that both the intermediates OH$^b$ and OH$^c$ oxidize to O. A key difference that we find in our analysis relative to that reported by Hansen et al. is that oxygen association at the cus sites ($2O^c \rightarrow O_2^{cc}$) is endothermic by 0.82 eV. This implies that $\mathrm{2O^c}$ is thermodynamically favorable relative to $O_2^{cc}$, which gets cascaded to the relative stability at higher potentials where we find that (i) $\mathrm{1O^c\ 1OOH^c}$ is more stable than $\mathrm{1HO^c\ 1O_2^{cc}}$, and (ii) $\mathrm{1O^c\ 1ClO^c}$ is more stable than $\mathrm{Cl(O^c)_2}$. This observation is comparable to the GGA-level DFT study of Exner et. al \cite{exner2014controlling} and Wang et al. \cite{wang2009intermediates}, who also predict that the $O_2^{cc}$ adsorbate on RuO$_2$ should dissociate into $2O^c$. We also observe that oxygen is the most stable adsorbate U$\approx$1.36 V; Cl does not adsorb directly on the cus site for chlorine evolution to occur through reaction pathway \ref{rxn3}. Therefore, we conclude from the phase diagram that chlorine evolution on RuO$_2$ could be mediated by reaction pathway \ref{rxn1}. It is worth highlighting that using error estimation capabilities we quantify the confidence in the prediction that the oxygen association at the cus site is endothermic, which is the origin of the differences we observe relative to Hansen et al. We find that greater than 99.9\% of the ensemble is in agreement with the prediction that oxygen association at the cus site is endothermic (refer to Figure S2 of the Supporting Information). This strengthens the need for c-value as a tool to quantify uncertainty in predictions of stable surface phases. The Pourbaix diagrams for the other rutile oxides are presented in the Supporting Information section.

\subsection{Scaling relations and generalized surface Pourbaix diagram}\label{SRGSPD}
The stable state of the surface $S$ is a function of $pH$, $U$ and the catalyst material $M_{cat}$. Pourbaix diagrams represent the stable surface state, $S(pH,U)$, on a specific material. However, to analyze trends in the predicted stable surface states across materials, a generalized Pourbaix diagram that represents $S(U,M_{cat})$ for a fixed $pH$ is required. This provides a computationally inexpensive way to understand the state of the catalyst surface without explicitly constructing Pourbaix diagrams for each material.

\begin{figure}[H]
	\centering
	\includegraphics[width=0.6\textwidth]{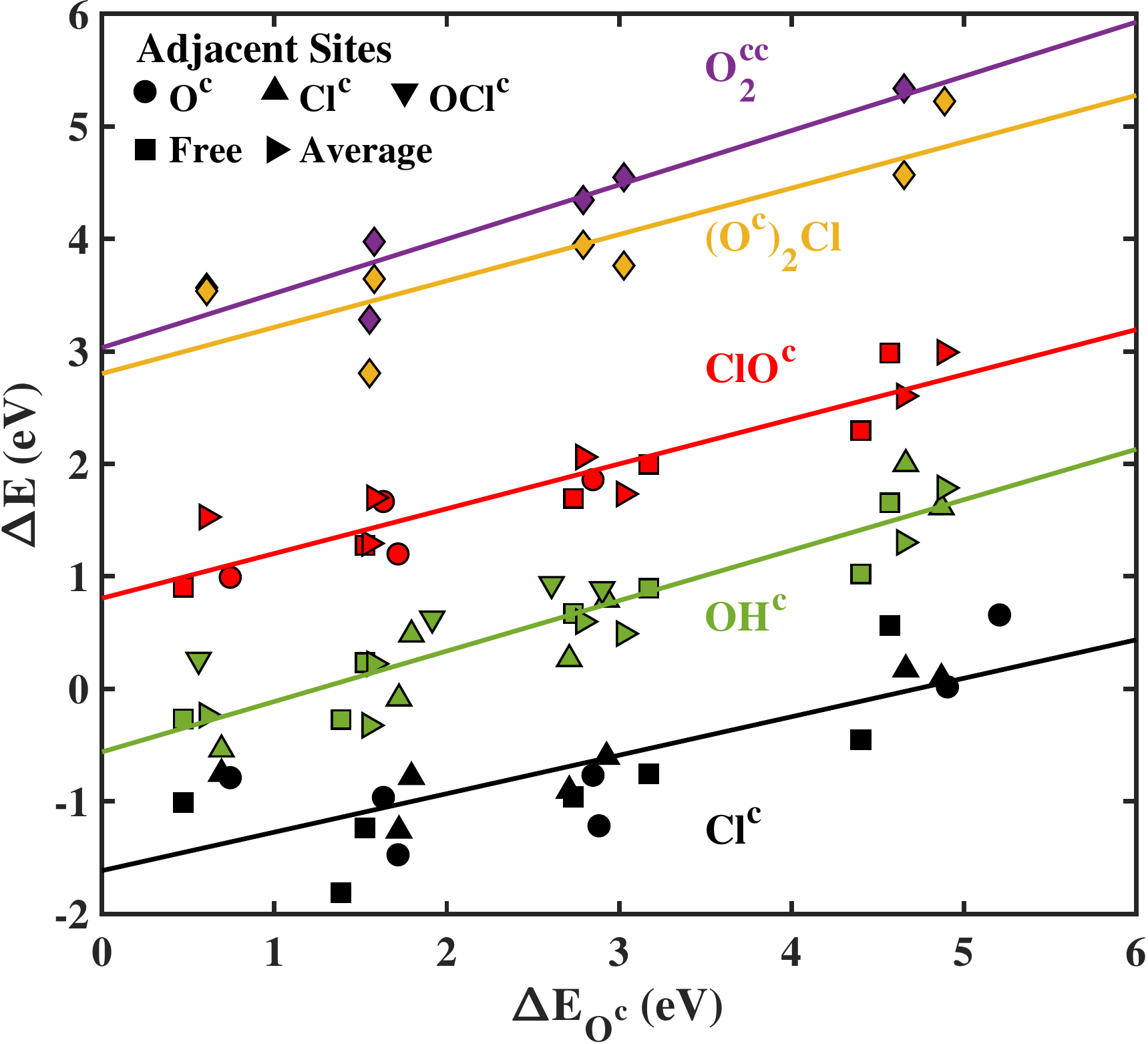}
	\caption{The adsorption free energies of various intermediates are plotted against the adsorption free energy of oxygen on the cus site of rutile oxides to show the scaling relations between various intermediates. The figure represents the adsorption energy of Cl (black) at the cus site: $\Delta E_{Cl^c} = 0.34 \Delta E_{O^c} - 1.62$ eV, adsorption energy of OH (green) at the cus site: $\Delta E_{HO^c} = 0.45 \Delta E_{O^c} - 0.56$  eV, adsorption energy of ClO (red) at the cus site: $\Delta E_{ClO^c} = 0.4 \Delta E_{O^c} + 0.81$, adsorption energy of $\mathrm{(O^c)_2Cl}$ (yellow) at the cus site: $\Delta E_{Cl(O^c)_2} = 0.41 \Delta E_{O^c} + 2.8$ eV and the adsorption energy of $\mathrm{O_2^{cc}}$ (purple) at the cus site: $\Delta E_{O_2^{cc}} = 0.48 \Delta E_{O^c} + 3.03$ eV. The markers represent the nature of the site neighboring the adsorption site: $\blacksquare$ - vacant neighboring cus-site, $\CIRCLE$ - O$\mathrm{^c}$ neighbor, $\blacktriangle$ - Cl$\mathrm{^c}$ neighbor, $\blacktriangledown$ - ClO$\mathrm{^c}$ neighbor. The marker \(\text{\Large$ \blacktriangleright $}\) represents the average adsorption energy of the intermediate species for the fully covered surface, and $\blacklozenge$ represents the adsorption of energy of $\mathrm{(O^c)_2Cl}$ (yellow) and $\mathrm{O_2^{cc}}$ (purple) plotted against the average adsorption energy of O for the fully covered surface. The fits here describe the scaling relations corresponding to the BEEF-vdW best-fit functional. The uncertainty in these fits is propagated through the outlined framework to quantify the confidence in the predicted generalized Pourbaix diagram (refer to Figure S4 in the supporting information).}
	\label{fig:All_Scalings}
\end{figure}

\begin{figure}[H]
	\centering
	\includegraphics[width=0.66\textwidth]{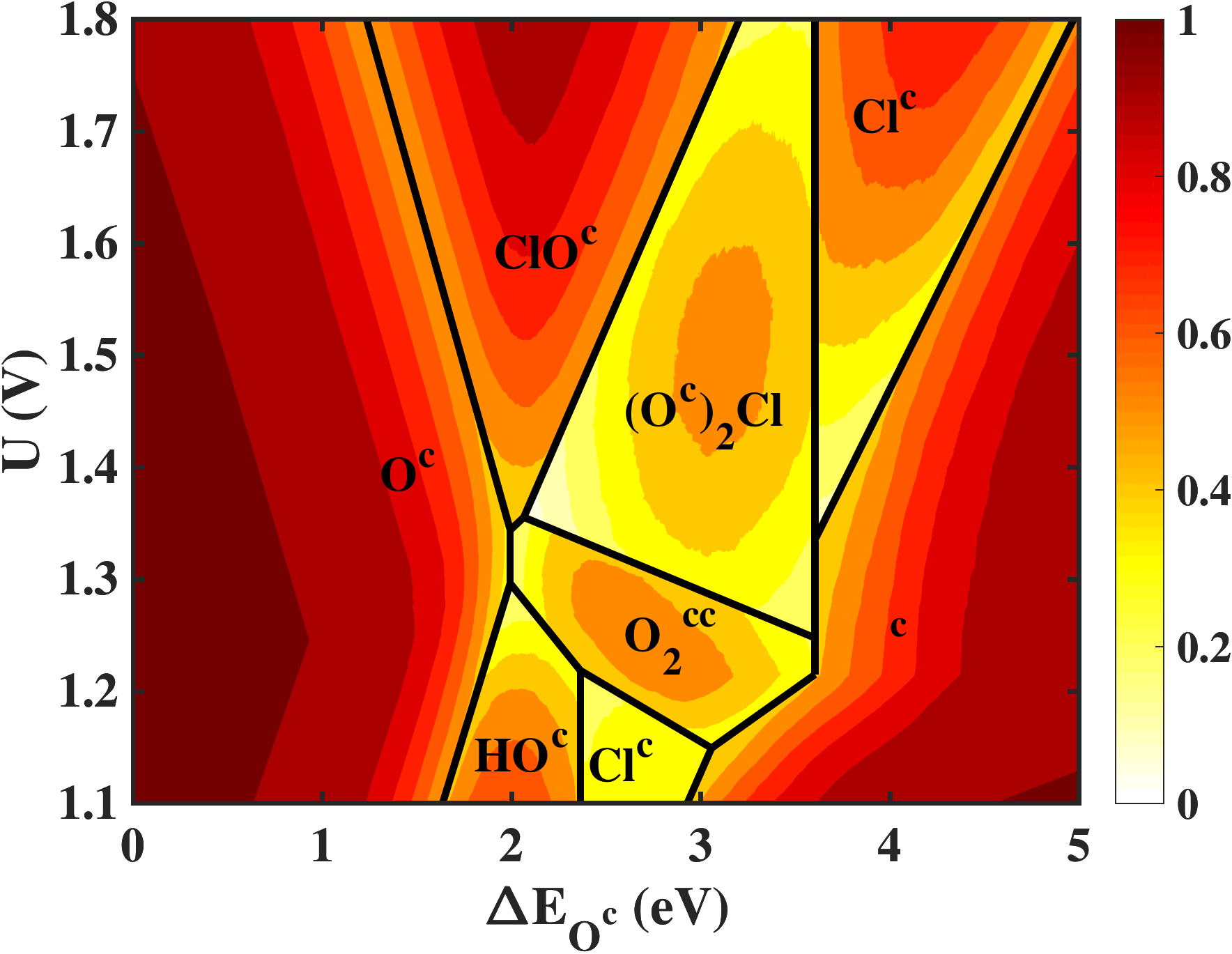}
	\caption{The generalized Pourbaix diagram showing the most stable surface at pH=0 and $\mathrm{a_{Cl^-}}$=1 as a function of potential and $\mathrm{\Delta E_{O^c}}$, the material descriptor. The black lines represent the phase boundaries of the generalized Pourbaix diagram constructed using the scaling relations obtained from the BEEF-vdW optimal functional. By propagating the uncertainty in the scaling relations, we quantify the confidence in the predictions as shown in the color-map. Regions of the plot with low c-value are correlated with the positions of surface phase boundaries, similar to that observed for the Pourbaix diagrams on individual rutile oxides.}
	\label{fig:Gen_p_cval}
\end{figure}

Generalized Pourbaix diagrams are constructed using a material descriptor to characterize the material $M_{cat}$, which is often chosen to be the adsorption energy of an intermediate. Hence, by invoking correlations (scaling relationships) between the adsorption energetics of various intermediates,~\cite{hansen2010electrochemical} we construct a single descriptor model to predict stable surface state across materials. The adsorption energies of various intermediates (Cl, OH, ClO, O$\mathrm{_2}$Cl and O$\mathrm{_2}$) at the cus site of the chosen oxides follow a linear relationship when plotted against the adsorption energy of oxygen at the cus site (using the BEEF-vdW optimal functional) as shown in Figure \ref{fig:All_Scalings}. This suggests that oxygen adsorption energy can be the continuous material descriptor choice for the reactivity of oxides. \cite{hansen2010electrochemical, pankratiev1982correlation} The rationale for this choice of material descriptor is the low sensitivity of adsorption energetics of other intermediates relative to $\Delta E_{O^c}$.  An alternate approach could be to choose the descriptor that maximizes prediction efficiency as shown in our earlier work.\cite{krishnamurthy2018maximal} For this work, we proceed with the choice of $O^c$ as the descriptor in order to perform comparison with Hansen \textit{et al.}.\cite{hansen2010electrochemical}  In this representation, we fix the $pH$ and Cl$^-$ concentration ($pH=0$, $a_{Cl^-}=1$), but their effects can be easily incorporated by changing the free energies of the various intermediate species accordingly. The species predicted in the generalized surface phase diagram as shown in Figure \ref{fig:Gen_p_cval} are consistent with those predicted by Hansen et al\cite{hansen2010electrochemical} although the positions of phase boundaries differ slightly. It is worth noting that in the weaker binding limit, i.e., $\Delta E_{O^c}>3.6$ eV (determined using the best-fit functional scaling relations), oxygen association at the cus site becomes exothermic leading to the reaction $O^{cc}_2 \rightarrow O_2(g) + 2^c$ becoming spontaneous.  Therefore the formation of the phases ClO$\mathrm{^c}$ and Cl$(\mathrm{O^c)_2}$ is unfavorable on catalyst materials with $\Delta E_{O^c}>3.6$ eV, and we only consider OH$\mathrm{^c}$, Cl$\mathrm{^c}$ and the clean surface as the possible stable surfaces states in this regime. 
In the single-descriptor picture, quantifying uncertainty in predictions becomes highly crucial since the uncertainty in DFT calculations is coupled with that in the scaling relations. We incorporate the uncertainty in scaling relations to quantify the confidence in predicted stable surface species. We generate an ensemble of predicted adsorption energies using the BEEF-vdW functional which results in an ensemble of scaling relations for each adsorbate. Therefore for each of the generated GGA-level XC, we can create a unique generalized Pourbaix diagram. We calculate the c-value of our predictions by finding the fraction of the functionals in agreement with the hypothesis corresponding to the BEEF-vdW best-fit functional at given value of $U$ and $\Delta E_{O^c}$ as described in section \ref{ss:cval_pourbaix}. We find that the c-value (Figure \ref{fig:Gen_p_cval}) of the predictions at the phase boundaries is the lowest, suggesting that precise determination of stable molecular states at these phase boundaries is computationally challenging. Lower c-values at phase boundaries originates from the uncertainty in scaling relationship(s) that are required to describe the phase boundary location. For instance, the phase boundary between the stable surface states $O^c$ and $ClO^c$ determined by $\Delta G(ClO^c)-\Delta G(O^c)-eU=0$ requires the scaling relation between $\Delta G(ClO^c)$ and $\Delta G(O^c)$ to be expressed in terms of the chosen material descriptor. This implies that the uncertainty in the position of this phase boundary stems from the uncertainty in the scaling relation. Similarly, the relatively higher uncertainty observed at the phase boundary between the stable surfaces states $ClO^c$ and $(O^c)_2Cl$ can be explained based on the fact that it requires propagation of uncertainty from two scaling relations involving $\Delta G(O^c)$. Section 6 of the supporting information provides a metric to quantify the likelihood of a certain surface state being the most stable by taking into account all of the members of the ensemble of functionals. 

For reducing uncertainty in predictions, surface phase diagrams with U and pH needs to be constructed for each given value of $\Delta E_{O^c}$ with the associated c-values. However, the identification of reaction mechanism trends is not possible with this approach. The generalized Pourbaix diagram with associated c-values constructed using the outlined approach maximizes identifiability of trends while retaining the ability to quantify uncertainty in the predictions. 

Our analysis suggests the general trend that as the $O^c$ binding strength on materials reduces ($0<\Delta E_{O^c}<5$), the stable surface state transitions from being completely covered with $O^c$ to a clean surface. Within this range, we limit our discussions to the chlorine containing species relevant for identifying the dominant reaction mechanisms.  For stronger binding materials, we observe that the $Cl^-$ anion is energetically more favorable as an adsorbed species on $O^c$ (as $ClO^c$) than on the cation of the clean surface. On relatively weaker binding materials, we  find that $Cl^-$ anion binds the surface as $(O^c)_2Cl$, which can be explained based on the stability of $(O^c)_2$ relative to 2$O^c$. For materials with weak binding ($\Delta E_{O^c}>3.6$ eV), oxygen can no longer adsorb on the cus site and $Cl^-$ directly adsorbs as $Cl^c$ on the catalyst.

 \subsection{Limiting potential and expected activity predictions}
 \label{LPEAP}
\begin{figure}[H]
	\centering
	\includegraphics[width=0.75\textwidth]{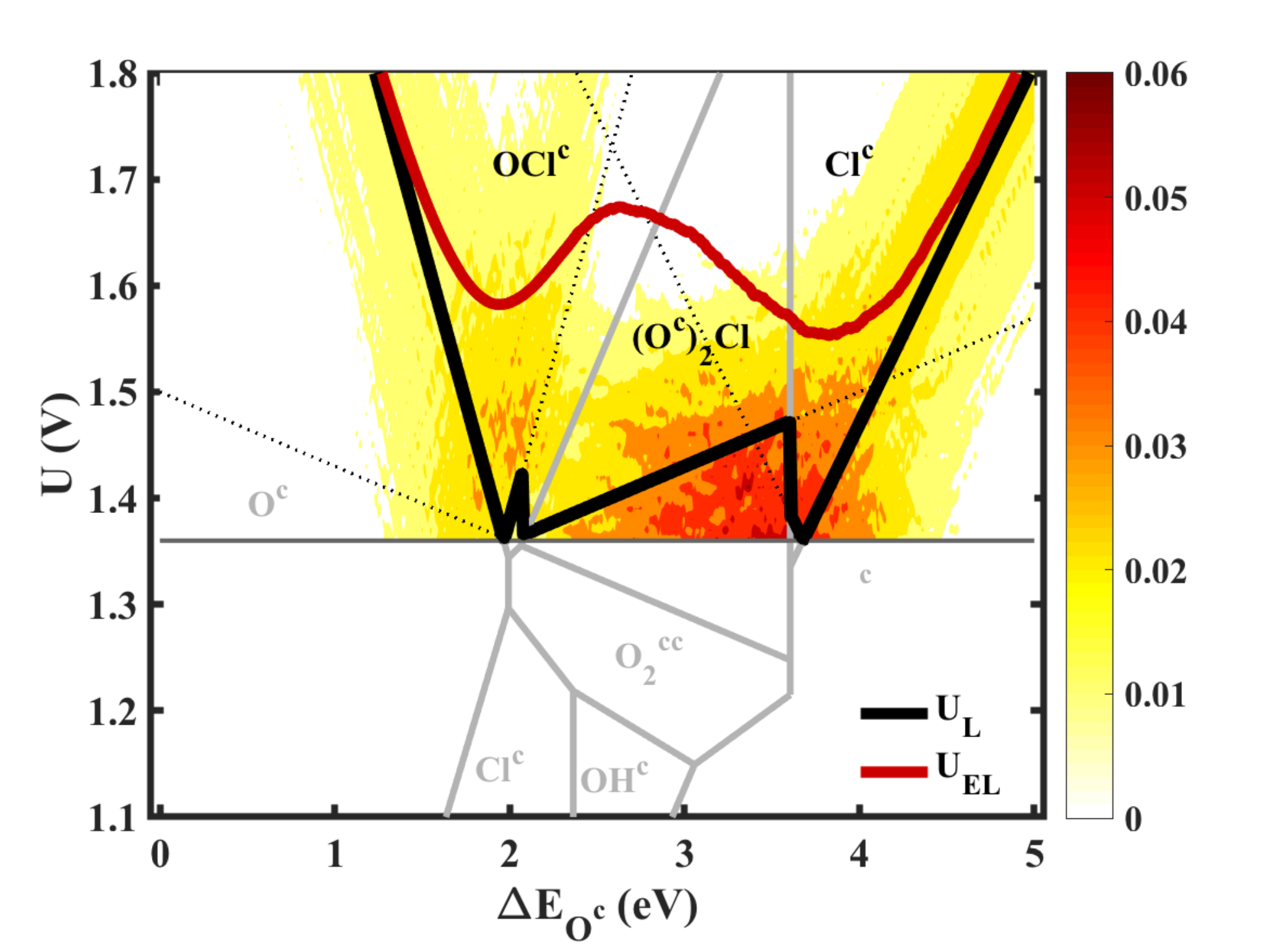}
	\caption{The activity volcano relationships (black dotted lines) for the possible reaction pathways involving ClO$^\mathrm{c}$, $\mathrm{(O^c)_2}$Cl and Cl$\mathrm{^c}$ (left to right) as a function of $\Delta E_{O^c}$ overlayed on the generalized Pourbaix diagram. The stable surface phases are labeled in gray and the gray lines indicate surface phase boundaries. The bold black lines show the generalized limiting potential relationship ($U_L$) as a function of $\Delta E_{O^c}$ and is constructed taking into account the stability of the underlying reaction intermediates using the BEEF-vdW optimal functional scaling relations. The color-map represents the probability distribution of the limiting potential quantified using the ensemble of functionals within the BEEF-vdW XC functional. The red line shows $U_{EL}(\Delta E_{O^c})$, the expectation value of the limiting potentials which is computed as the probability weighted average of the $U_L$ distribution at each descriptor value. We observe a large deviation of $U_{EL}$ from the $U_L$ for $\sim$2.1<$\Delta E_{O^c}$<3.6 eV, where the reaction mechanism mediated by $\mathrm{(O^c)_2Cl}$ is predicted to be active for chlorine evolution. We attribute this difference to the fact that the construction of the activity volcano for this reaction pathway involves two scaling relations, which increases the uncertainty in the predictions.}
	\label{fig:UEL}
\end{figure}

 Computational screening for active materials requires the understanding of trends in catalytic activity as a function of the chosen material descriptor. For a given material, the activity is governed by the operative reaction mechanism, which is dictated by the stable intermediate species on the surface given by the generalized surface Pourbaix diagram (Figure \ref{fig:Gen_p_cval}). Invoking linear scaling relationships between the adsorption free energy of oxygen and the adsorption free energies of all the other intermediates allows us to represent both the limiting potential for each mechanism and the stable phase space (generalized Pourbaix diagram) as a function of the same variables -- $U$ and $\Delta E_{O^c}$ (Figure \ref{fig:UEL}). Overlaying the generalized surface phase diagram with the activity volcano is crucial to obtain the generalized activity volcano (Figure \ref{fig:UEL}) by accounting for the appropriate active reaction mechanism and active site. This results in the activity being governed by reaction mechanism \ref{rxn1} for strong binding materials ($\Delta E_{O^c} < \lesssim 2.1 (eV)$), by reaction mechanism \ref{rxn1} for moderate binding materials ($ 2.1 \lesssim \Delta E_{O^c} \lesssim 3.6$ eV), and by reaction mechanism \ref{rxn3} for weak binding materials ($\Delta E_{O^c} \gtrsim 3.6 $eV). 
 
Such single-descriptor activity-prediction models have been successful in determining promising catalysts for various electrochemical reactions including hydrogen evolution\cite{greeley2006computational, greeley2006hydrogen, norskov2005trends}, oxygen reduction\cite{greeley2009combinatorial,viswanathan2012universality}, hydrogen peroxide synthesis,\cite{rankin2012trends,viswanathan2015selective,verdaguer2014trends} and oxygen evolution \cite{man2011universality, halck2014beyond}. Alongside, we address an associated question of the level of confidence in activity predictions, and identify material descriptor regimes where the prediction uncertainty is low. We incorporate the uncertainty in scaling relationships and use a Bayesian error estimation approach using the BEEF-vdW XC to quantify the confidence in the predictions of reaction mechanisms and thereby the activity for chlorine evolution. The XC functional generates an ensemble of adsorption energies, as shown in Figure S1, that results in an ensemble of scaling relations for each of the intermediates. Hence, each GGA-level XC functional generated within the ensemble leads to a unique generalized Pourbaix diagram and an associated generalized activity volcano. At a given $\Delta E_{O^c}$, the specific reaction mechanism active for chlorine evolution which governs the limiting potential can hence be determined by the associated stable intermediate species through the generalized Pourbaix diagram. Therefore, for a given value of the material descriptor, $\Delta E_{O^c}$, we can now find a distribution of the predicted limiting potentials corresponding to the family of functionals. This can be used to construct a probability map of the limiting potential for a range of descriptor values, as can be seen in Figure \ref{fig:UEL}. We report $U_{EL}(\Delta E_{O^c})$, the expectation value of the limiting potential, which is computed as the probability weighted average (refer to section \ref{ss:UEL}) of the $U_L(\Delta E_{O^c})$.~\cite{krishnamurthy2018maximal,deshpande2016quantifying} 


We observe that the $U_{EL}$ curve overlaps with the $U_L$ away from the peaks of the volcanoes owing to low prediction uncertainty.  Similar observations have been made for hydrogen evolution, oxygen reduction and oxygen evolution reaction.\cite{krishnamurthy2018maximal} It is worth pointing out that the $U_{EL}$ predictions based on the activity volcano for the reaction pathway mediated by $\mathrm{Cl(O^c)_2}$ and $\mathrm{O^{cc}_2}$ are relatively much higher than the $U_L$. This can be rationalized partly based on the fact that the activity relationship for this pathway alone involves uncertainty incorporation from two scaling relations: ($\Delta E_{O_2^{cc}}$ and $\Delta E_{Cl(O^c)_2}$ as functions of $\Delta E_{O^c}$).
A similar observation was reported for predicted activity of transition metals for the oxygen reduction reaction using $\Delta E_{O^*}$ as the descriptor.~\cite{krishnamurthy2018maximal} An additional factor that lowers the prediction confidence in this regime is the low associated c-values ($<\approx 0.55$) for regions in the generalized surface Pourbaix diagram where $\mathrm{Cl(O^c)_2}$  and $\mathrm{O^{cc}_2}$ are predicted to be the most stable surface state as can be in Figure \ref{fig:Gen_p_cval}. Hence, DFT predictions of chlorine evolution activity for $\Delta E_{O^c}$ approximately in the range 2.1-3.6 eV are subject to higher uncertainty. We argue that $U_{EL}(\Delta E_{O^c})$ is a more relevant activity measure since it explicitly incorporates DFT uncertainty.

\subsection{Quantifying confidence in the predicted active reaction pathway}
In reaction schemes involving multiple pathways, an equally important question involves identifying the dominant reaction pathway as a function of the material descriptor.  Here, we address the question of propagating the uncertainty associated with adsorption energies and scaling relations to quantify the confidence in predicted reaction pathway.  This allows us to predict the active reaction mechanism for chlorine evolution for any new catalyst surface, to a first approximation, through the adsorption free energy of oxygen on the oxide ($\Delta E_{O^c}$) as the descriptor. We determine the confidence in the predicted reaction mechanism through the metric defined earlier in section \ref{ss:cval_rxn_mech}, $c_{m_{pred}}$, as a function of $\Delta E_{O^c}$. $c_{m_{pred}}(\Delta E_{O^c})$ is calculated based on the level of agreement between the predicted reaction mechanism(s) by the ensemble of GGA-level functionals and the BEEF-vdW optimal functional. 
We define another useful quantity, $c_{m_{pred}=i}$ for the $i^{\mathrm{th}}$ reaction mechanism, which becomes especially important in regimes where the $c_{m_{pred}}$ value is less than 1, where different functionals identify different reaction mechanisms to be active. $c_{m_{pred}=i}(\Delta E_{O^c})$ provides a measure of the confidence for the $i^{\mathrm{th}}$ reaction mechanism being active (refer to eqn. \ref{cmechi}).
\begin{figure}[H]
	\centering
	\includegraphics[width=0.6\textwidth]{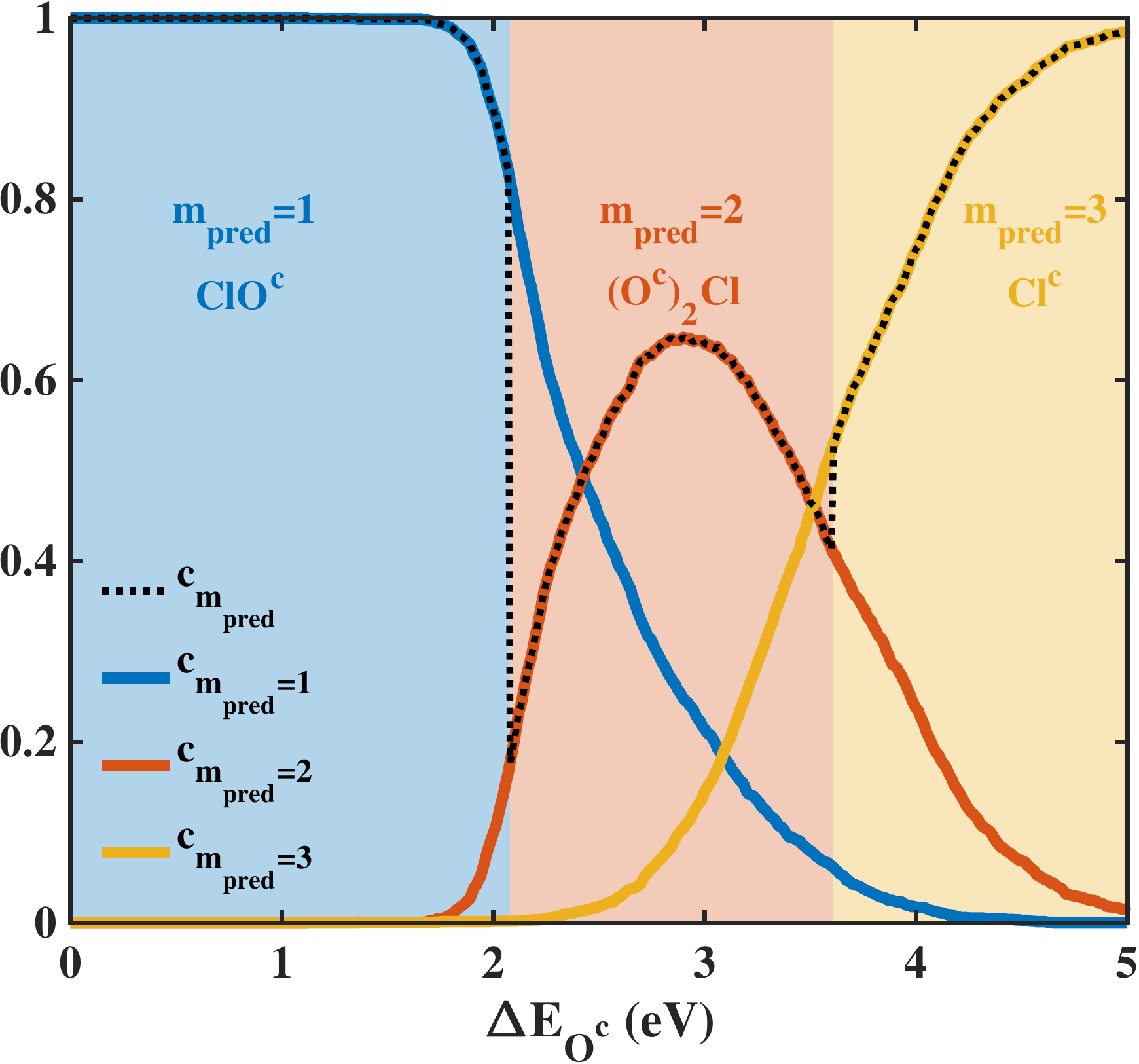}
	\caption{Quantifying the confidence in the predicted reaction mechanism $c_{m_{pred}}$ as a function of the material descriptor (Black dotted). The bold lines represent the confidence in the reaction mechanisms mediated by the stable intermediates $\mathrm{ClO^c}$ ($c_{m_{pred}=1}$), $\mathrm{(O^c)_2Cl}$ ($c_{m_{pred}=2})$ and $\mathrm{Cl^c}$ ($c_{m_{pred}=3}$). The colored regions in the plot represent the reaction mechanism predicted by the BEEF-vdW (best-fit) optimal functional. We observe relatively low $c_{m_{pred}}$ for reaction mechanism mediated by intermediate $\mathrm{(O^c)_2Cl}$ in the material descriptor range $\sim$ 2.1<$\Delta E_{O^c}$<3.6 eV implying a higher degree of disagreement between the GGA-level functionals within the ensemble of functionals.}
	\label{fig:cval_mech}
\end{figure}

We observe (Figure \ref{fig:cval_mech}) that for materials with strong oxygen binding ($\approx 0 <\Delta E_{O^c}< 2.1$ eV), the majority of functionals predict that chlorine evolution occurs by the reaction mechanism mediated by ClO$\mathrm{^c}$ ($c_{m_{pred}} \approx c_{m_{pred}=1} \approx 1)$. In the range of materials that moderately bind oxygen on the cus cite ($\approx 2.1< \Delta E_{O^c} <\approx 3.6$ eV), we observe that there exists a lower degree of agreement between the functionals with respect to the predicted reaction mechanisms leading to relatively low $c_{m_{pred}}$ values although most functionals are in agreement with the optimal around $\Delta E_{O^c}\approx 3 eV$. Materials that bind oxygen on the cus site weakly ($\Delta E_{O^c} > 3.6$ eV), it is thermodynamically favorable to have the chlorine intermediate directly adsorbed on the cus site rather than on oxygen (as either $\mathrm{ClO^c}$ or $(O^c_2)Cl$). In this regime, we find that majority of functionals agree with the BEEF-vdW optimal functional, and chlorine evolution is mediated with a high certainty by the reaction pathway mediated by Cl$\mathrm{^c}$. 

It is worth noting that different functionals may identify the same reaction pathway as the dominant one.  However, they may result in vastly different limiting potentials. This raises an important question on the confidence associated with identifying reaction mechanisms versus determining activity values.  This can be analyzed through an example case of RuO$_2$ whose $\Delta E_{O^c} = 1.53~eV$.   In this case, all the functionals identify the $\mathrm{ClO^c}$ mediated pathway to be the active one.  However, the functionals identify vastly different limiting potentials, as shown in Fig S7.  This leads to an important conclusion that the confidence in utilizing DFT calculations for the classification problem of identifying the dominant reaction mechanism is much better than prediction problem of determining the catalytic activity.  This has an important implication more broadly that DFT-identified reaction pathways are likely to be much more robust than the DFT-predicted catalytic activities.

\section{Conclusions} \label{Conclusions}
We demonstrate an approach to quantify the confidence associated with predictions of the activity-governing reaction mechanism by systematically accounting for the uncertainty derived from the DFT-calculated adsorption energetics. By using c-value in the context of Pourbaix diagrams as a metric to quantify the level of confidence in surface state predictions, we highlight and rationalize the origin of higher uncertainty close to phase boundaries. We quantitatively suggest with high confidence that chlorine evolution on RuO$_2$ and IrO$_2$ occurs through the reaction mechanism mediated by ClO$^c$. To analyze trends in the predicted stable surface states across materials, we construct a generalized Pourbaix diagram by invoking scaling relations between the surface intermediate species. We choose $\Delta E_{O^c}$ as the material descriptor since it allows greater resolution of material candidates on the descriptor scale. By overlaying the generalized surface phase diagram with the activity relationships we construct the generalized activity volcano by accounting for the appropriate operative reaction mechanism. Incorporating the uncertainty in the scaling relations, we report the prediction confidence in the generalized Pourbaix diagram and find the distribution of the limiting potential as a function of the material descriptor. We compute the expected activity, $U_{EL}(\Delta E_{O^c})$, which we argue is a more relevant activity measure since it incorporates DFT uncertainty. Furthermore, we quantify the prediction robustness of the likely operative reaction mechanism as a function of the chosen material descriptor, which we believe is crucial for improving the accuracy of activity predictions in descriptor-based catalyst screening approaches. Although this work focuses on the mechanisms by which chlorine evolution can occur, the presented approach can readily be extended not only to other multi-electron reactions involving multiple reaction pathways but also to reactions where selectivity of products is critical.    

\section*{Acknowledgements}\label{sec:ack}
The authors thank Dr. Heine A. Hansen for insightful discussions and for sharing structure files. D.K. and V.V gratefully acknowledge funding support from the National Science Foundation under award CBET-1554273.  V. S. and V. V. acknowledge support from the Scott Institute for Energy Innovation at Carnegie Mellon University. \\

\noindent \textbf{Supporting Information Available:} Computational details, adsorption energetics of all the considered surface states, scaling relationships, Pourbaix diagram construction, uncertainty propagation framework for chlorine evolution activity.

\bibliographystyle{achemso}
\bibliography{refs}


\section*{Supplementary material}

# Supporting Information for Quantifying Confidence in DFT Predicted Surface Pourbaix Diagrams and the Associated Reaction Pathways for Chlorine Evolution

Vaidish Sumaria,[†] Dilip Krishnamurthy,[‡] and Venkatasubramanian Viswanathan[*,‡,†]

[†]Department of Chemical Engineering, Carnegie Mellon University, Pittsburgh, PA 15213, USA

[‡]Department of Mechanical Engineering, Carnegie Mellon University, Pittsburgh, PA 15213, USA

# 1 Adsorption Energy Calculation

We consider the adsorporption of $O^c$, $HO^c$, $HOO^c$, $Cl^c$, $ClO^c$, $O_2^{cc}$ and $Cl(O^c)_2$ and their combinations on the cus site and $OH^b$ on the bridge site. The adsorption energies (relative to clean catalysts surface) of there intermediated were calculated using the following equations:

$$\Delta E_{O^c} = E_{O^c} - E_c - E_{H_2O} + E_{H_2} \tag{1a}$$

$$\Delta E_{OH^c} = E_{OH^c} - E_c - E_{H_2O} + 1/2 E_{H_2} \tag{1b}$$

$$\Delta E_{OOH^c} = E_{OOH^c} - E_c - 2E_{H_2O} + 3/2 E_{H_2} \tag{1c}$$

$$\Delta E_{Cl^c} = E_{Cl^c} - E_c - 1/2 E_{Cl_2} \tag{1d}$$

$$\Delta E_{ClO^c} = E_{ClO^c} - E_c - 1/2 E_{Cl_2} - E_{H_2O} + E_{H_2} \tag{1e}$$

$$\Delta E_{O_2^{cc}} = E_{O_2^{cc}} - E_{2c} - 2E_{H_2O} + 2E_{H_2} \tag{1f}$$

$$\Delta E_{Cl(O^c)_2} = E_{Cl(O^c)_2} - E_{2c} - 2E_{H_2O} + 2E_{H_2} - 1/2 E_{Cl_2} \tag{1g}$$

$$\Delta E_{OH^b} = E_{OH^b} - E_{O^b} - 1/2 E_{H_2} \tag{1h}$$

The adsorption energies of these intermediates calculated using the BEEF-vdW best-fit functional are reported in Table S1.

# 2 Adsorption Energy Distribution

Using the BEEF-vdW functional, ensemble of adsorption energies for various intermediates involved is generated. The Figure S1 shows the combined distribution of adsorption free energy for $IrO_2$, $TiO_2$, $PtO_2$, $RhO_2$, $SnO_2$ and $VO_2$ centered around $\Delta E_{O^c} = 0$ eV.

Table S1: Adsorption Energies of the various intermediates considered on the surface of rutile (110) surface. All the energies reported are in eV. Entries in the table shown as "-" represent that a reconstructed surface is thermodynamically more favorable.

| | $RuO_2$ | $IrO_2$ | $TiO_2$ | $PtO_2$ | $RhO_2$ | $VO_2$ | $SnO_2$ |
|---|---|---|---|---|---|---|---|
| $2HO^b\ 2HO^c$ | -2.02 | -2.58 | -1.02 | -2.81 | -2.62 | | -2.8 |
| $2HO^b$ | -1.97 | -1.45 | 0.1 | -2.92 | -3.33 | | -1.26 |
| $2O^b\ 2HO^c$ | 0.44 | -0.65 | 3.57 | 0.98 | 1.19 | -0.47 | 2.6 |
| $2O^b\ 1HO^c\ 1Cl^c$ | -0.75 | -1.9 | 2.56 | -0.49 | -0.17 | -1.55 | 1.16 |
| $2O^b\ 1Cl^c$ | -1.23 | -1.81 | 0.56 | -0.76 | -0.96 | -1.01 | -0.45 |
| $2O^b\ 2Cl^c$ | -2.02 | -3.07 | 0.74 | -1.66 | -1.56 | -1.77 | -0.36 |
| $2O^b\ 1HO^c$ | 0.23 | -0.27 | 1.65 | 0.89 | 0.67 | -0.27 | 1.02 |
| $2O^b\ 1O^c\ 1Cl^c$ | 0.56 | -0.09 | 5.23 | 1.95 | 1.97 | -0.32 | 4.42 |
| $2O^b\ 1O^c$ | 1.53 | 1.39 | 4.57 | 3.17 | 2.74 | 0.47 | 4.4 |
| $2O^b\ 1HO^c\ 1OCl^c$ | 1.9 | 0.91 | 4.85 | 2.18 | 2.58 | 1.16 | - |
| $2O^b\ 1HOO^c$ | 3.04 | 2.64 | 4.62 | 3.48 | 3.46 | 2.9 | 4.1 |
| $2O^b\ 2O^c$ | 3.16 | 3.1 | 9.78 | 6.06 | 5.58 | 1.21 | 9.31 |
| $2O^b\ 1O_2^{cc}$ | 3.97 | 3.28 | - | - | 4.35 | 3.57 | 5.34 |
| $2O^b\ 1O^c\ 1ClO^c$ | 3.19 | 2.59 | - | - | 4.59 | 1.46 | - |
| $2O^b\ 1Cl(O^c)_2$ | 3.64 | 2.81 | 5.22 | 3.76 | 3.95 | 3.54 | 4.57 |
| $2O^b\ 1O^c\ 1HOO^c$ | 4.86 | 4.27 | - | - | 6.35 | - | - |
| $2O^b\ 2ClO^c$ | 3.4 | 2.59 | 5.98 | 3.46 | 4.12 | 3.06 | 5.2 |
| $2O^b\ 1ClO^c$ | 1.28 | - | 2.98 | 2 | 1.69 | 0.9 | 2.29 |

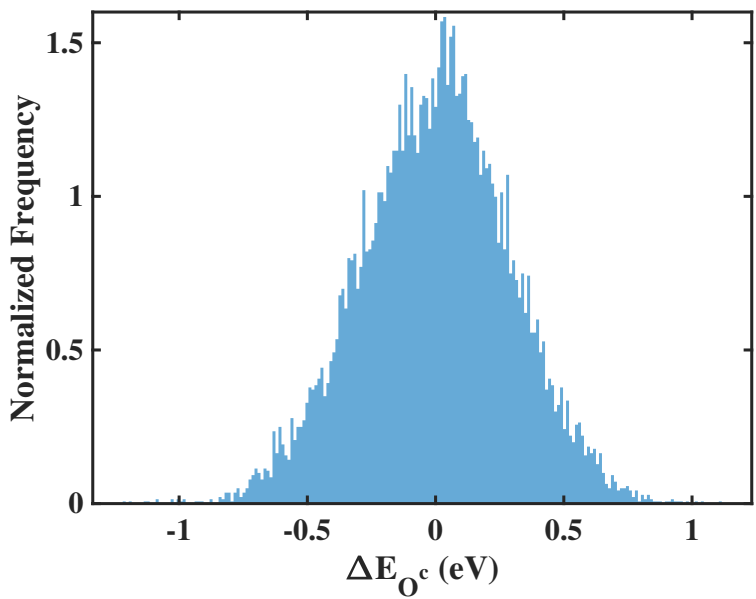

Figure S1: Combined distribution of oxygen adsorption free energies on the cus-site for $IrO_2$, $RuO_2$, $TiO_2$, $PtO_2$, $RhO_2$, $SnO_2$ and $VO_2$, re-centered around $\Delta E_{O^c} = 0$ eV.

# 3 Zero Point and Entropy Corrections

We report the $\Delta ZPE$ and $\Delta S°$ for all the relevant reactions at T=298K in Table S2.

Table S2: Zero point and entropic corrections at 298 K for rutile oxides (110)

|  | TS | T$\Delta$S | ZPE | $\Delta$ZPE | $\Delta ZPE - T\Delta S$ |
|---|---|---|---|---|---|
| $H_2O(l)$ | 0.67 | 0 | 0.56 | 0 | 0 |
| $H_2O \rightarrow HO^c + 1/2H_2$ | 0.20 | -0.47 | 0.50 | -0.06 | 0.41 |
| $H_2O \rightarrow O^c + H_2$ | 0.41 | -0.27 | 0.34 | -0.22 | 0.05 |
| $1/2Cl_2 \rightarrow Cl^c$ |  | -0.34 |  | -0.02 | 0.37 |
| $H_2$ | 0.41 |  | 0.27 |  |  |
| $1/2O_2$ | 0.32 |  | 0.05 |  |  |
| $Cl_2$ | 0.69 |  | 0.06 |  |  |
| $O^c$ | 0 |  | 0.07 |  |  |
| $OH^c$ | 0 |  | 0.36 |  |  |
| $H^c$ | 0 |  | 0.17 |  |  |

# 4 Pourbaix Diagram Construction

For the construction of the surface Pourbaix Diagrams, we calculate the free energy of adsorbates as a function of potential (U) and pH. These adsorbates are formed by water discharge ($O^c, HO^c, HOO^c$) and adsorption of chloride ions.

$$H_2O(l) + {}^c \rightarrow O^c + 2H^+(aq) + 2e^- \tag{2a}$$

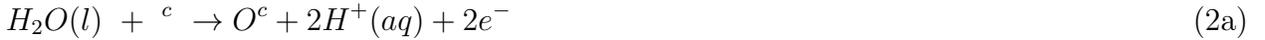

$$H_2O(l) + {}^c \rightarrow HO^c + H^+(aq) + e^- \tag{2b}$$

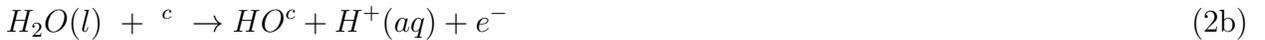

$$2H_2O(l) + {}^c \rightarrow HOO^c + 3H^+(aq) + 3e^- \tag{2c}$$

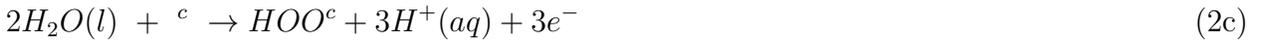

$$Cl^-(aq) + {}^c \rightarrow Cl^c + e^- \tag{2d}$$

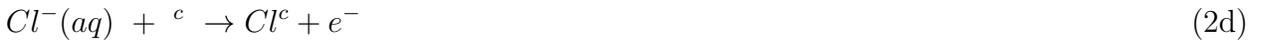

The free energy of $O^c$ as defined by eq. 2a can be expressed as

$$\Delta G_{O^c}(U, a_{H^+}) = G_{O^c} + 2\mu_{H^+} + 2\mu_{e^-} - \mu_{H_2O(l)} - G_c \tag{3}$$

4

The chemical potential of protons and electrons can be expressed as:

$$\mu_{H^+} = \mu^\circ_{H^+} + k_B T ln(a_{H^+}) \qquad (4)$$

$$\mu_{e^-} = \mu^\circ_{e^-} - eU \qquad (5)$$

where, $\mu^\circ_{H^+}$ is the chemical potential of proton at standard conditions ($p_{H_2} = 1$ bar, $pH = 0$) and $\mu^\circ_{e^-}$ is an arbitrary reference energy determined by the potential scale. We reference the potential relative to the standard hydrogen electrode (SHE). Hence, the hydrogen electrode reaction,

$$H^+(aq) + e^- \leftrightarrow 1/2\ H_2 \qquad (6)$$

is at equilibrium at zeo potential and standard conditions. This gives:

$$\mu^\circ_{H^+} + \mu^\circ_{e^-} = \mu^\circ_{H_2} \qquad (7)$$

The chemical potential of liquid water is calculated from water is calculated from water vapor in equilibrium with liquid water at 298 K and 0.035 bar. Hence, using eq. 4, 5, 7 and 3, the formation energy of $O^c$ can be expressed as:

$$\Delta G_{O^c}(U, a_{H^+}) = G_{O^c} - G_c - \mu_{H_2O} + \mu^\circ_{H_2} + 2(k_B T ln(a_{H^+}) - eU)$$

$$= \Delta E_{O^c} + \Delta ZPE - T\Delta S^\circ + 2(k_B T ln(a_{H^+}) - eU)$$

$$= \Delta G^\circ_{O^c} + 2(k_B T ln(a_{H^+}) - eU) \qquad (8)$$

Where, $\Delta E_{O^c}$ is given by eq. 1a, $\Delta ZPE$ is the change in zero point energy upon adsorption and $\Delta S$ includes the loss of translational entropy of the gas phase molecules upon adsorption on the surface. The change in zero point energy upon adsorption is calculated within the harmonic approximation for an adsorbate at the cus-site of $RuO_2$ and is assumed to be constant for all rutile oxides.[1]

The free energy of adsorption of chlorine as defined by eq. 2d can be expressed as

$$\Delta G_{Cl^c}(U, a_{Cl^-}) = G_{Cl^-} - G_c + \mu_{e^-} - \mu^\circ_{Cl^-} - k_B T ln(a_{Cl^-}) \qquad (9)$$

We use the standard reversible chlorine electrode to calculate the chemical potential of chlorine ion. The chlorine electrode is in equilibrium at standard conditions, and can be expressed as,

$$Cl^-(aq) \leftrightarrow 1/2\ Cl_2(g) + e^- \qquad (10)$$

$$\mu°_{Cl-} = 1/2\, \mu°_{Cl_2} + \mu°_{e-} - eU°_{Cl-} \qquad (11)$$

Where $U°_{Cl-}$ is the standard potential of the reversible chlorine and $U°_{Cl-} = 1.36$ V at 298K. Hence, substituting eq. 11 in eq. 9, we can express the free energy of adsorption of chlorine as,

$$\begin{aligned}
\Delta G_{Cl^c}(U, a_{Cl-}) &= G_{Cl^c} - G_c + 1/2\, \mu°_{Cl_2} - e(U - U°_{Cl}) - k_B T ln(a_{Cl-}) \\
&= \Delta E_{Cl^c} + \Delta ZPE - T\Delta S° - k_B T ln(a_{Cl-}) - e(U - U°_{Cl}) \\
&= \Delta G°_{Cl^c} - k_B T ln(a_{Cl-}) - e(U - U°_{Cl}) \qquad (12)
\end{aligned}$$

In this work, we consider $a_{Cl-}=1$. Hence, the free energy of adsorption is only a function of $U$ and $a_{H+}$. Although, the effect of activity of chlorine ions can be treated just by changing the free energies accordingly.

In order to find the free energy of adsorption, as a function of $U$ and $pH$, we use the following methodology:

$$k_B T ln(a_{H+}) = ln(10)\, k_B T\, \log_{10}(a_{H+}) = -0.059\, pH(eV) \qquad (13)$$

Hence, the free energy of any adsorbate for $a_{Cl-}=1$ can be rewritten in the form:

$$\boldsymbol{\Delta G(U, pH) = \Delta G° - 0.059\, \nu_{H+} \times pH - \nu_{e-}\, eU - \nu_{Cl-}\, eU_{Cl}} \qquad (14)$$

Where, $\nu_{H+}$, $\nu_{e-}$ and $\nu_{Cl-}$ is the number of protons, electrons and chlorine ions involved in the reaction defining the adsorption of a given adsorbate species and are listed in the Table S3.

Table S3: List of coefficients: $\nu_{H^+}$, $\nu_{e^-}$ and $\nu_{Cl^-}$ for the relevant adsorbate structures required to calculate the free energy using the eq. 14

| Structure | $\nu_{H^+}$ | $\nu_{Cl^-}$ | $\nu_{e^-}$ |
|---|---|---|---|
| 2HO$^b$ 2HO$^c$ | 0 | 0 | 0 |
| 2HO$^b$ | -2 | 0 | -2 |
| 2O$^b$ 2HO$^c$ | 2 | 0 | 2 |
| 2O$^b$ 1HO$^c$ 1Cl$^c$ | 1 | -1 | 2 |
| 2O$^b$ 1Cl$^c$ | 0 | -1 | 1 |
| 2O$^b$ 2Cl$^c$ | 0 | -2 | 2 |
| 2O$^b$ 1HO$^c$ | 1 | 0 | 1 |
| 2O$^b$ 1O$^c$ 1Cl$^c$ | 2 | -1 | 3 |
| 2O$^b$ 1O$^c$ | 2 | 0 | 2 |
| 2O$^b$ 1HO$^c$ 1ClO$^c$ | 3 | -1 | 4 |
| 2O$^b$ 1HOO$^c$ | 3 | 0 | 3 |
| 2O$^b$ 2O$^c$ | 4 | 0 | 4 |
| 2O$^b$ 1O$^{cc}_2$ | 4 | 0 | 4 |
| 2O$^b$ 1O$^c$ 1ClO$^c$ | 4 | -1 | 5 |
| 2O$^b$ 1Cl(O$^c$)$_2$ | 4 | -1 | 5 |
| 2O$^b$ 1O$^c$ 1HOO$^c$ | 5 | 0 | 5 |
| 2O$^b$ 2ClO$^c$ | 4 | -2 | 6 |
| 2O$^b$ 1ClO$^c$ | 2 | -1 | 3 |

Hence using eq. 14 and Table S3, we find the free energy of adsorption of all the adsorbates at any given potential and $pH$. The Pourbaix diagram depicts the adsorbate species that has the least free energy of formation in the potential and $pH$ space.

**RuO$_2$ Pourbaix diagram**

The Figure S2, shows the distribution of the energy change for oxygen association at the cus sites. We observe that greater than 99.9% of the ensemble predicts that oxygen association at the cus sites ($2O^c \rightarrow O^{cc}_2$) is endothermic as can be seen in the distribution shown in Figure S2.

**Pourbaix diagrams of the other Rutile oxides**

The Pourbaix diagrams for PtO$_2$, TiO$_2$, RhO$_2$ and SnO$_2$ are shown in the Figures S3a, S3b, S3c and S3d respectively. In all the cases, we make a similar observation that the c-value is low at the phase boundaries, hence stable phase predictions in those regimes is highly uncertain.

7

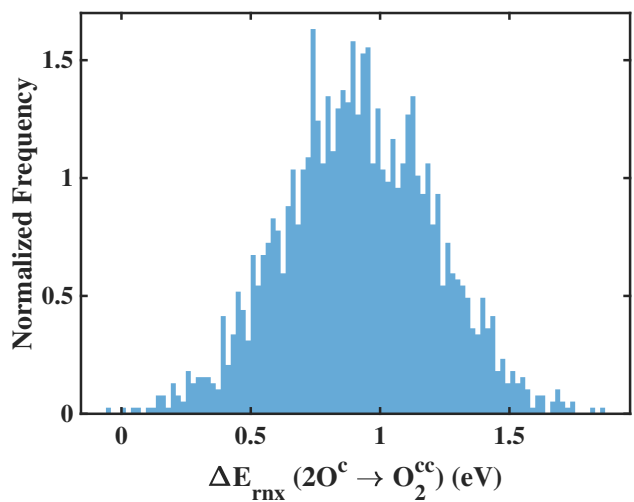

Figure S2: Distribution of the reaction energy for association of oxygen on the cus site.

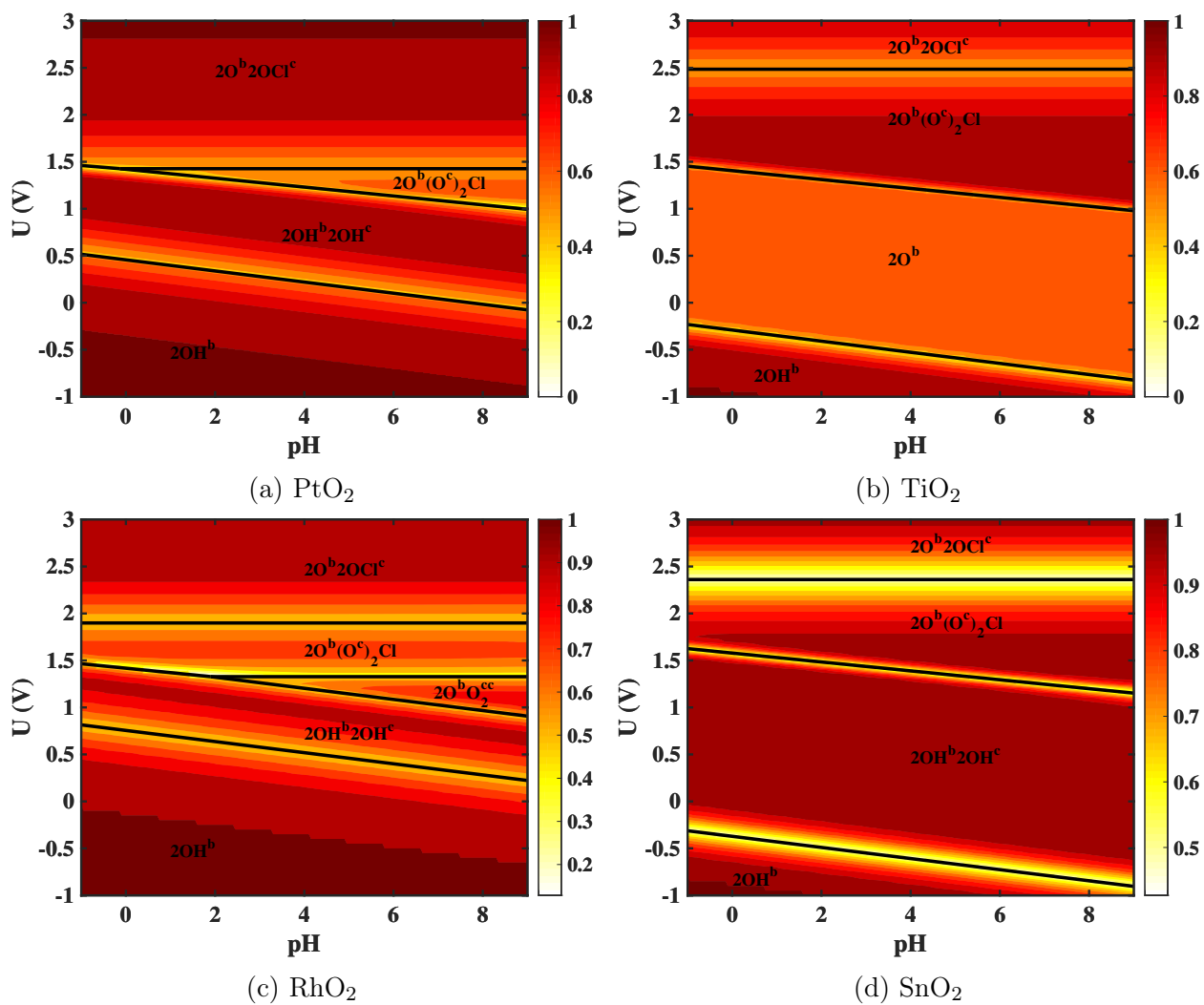

Figure S3: Pourbaix diagrams for various rutile (110) oxide surfaces with the associated c-value contours. The c-values are computed to be low for electrochemical regimes that represent the stable surface phase boundaries owing to the uncertainty in scaling relationship(s).

8

# 5 Scaling Relations, Generalized Pourbaix Diagram and Activity Volcano

To construct the generalized pourbaix diagram and the activity volcano, we use the scaling relationships. Eq. 15 defines the linear fits of the BEEF-vdW best-fit functional for the various reaction intermediates.

$$\Delta E_{Cl^c} = 0.34 \Delta E_{O^c} - 1.62 \ eV; \quad MAE = 0.32 \ eV \tag{15a}$$

$$\Delta E_{ClO^c} = 0.4 \Delta E_{O^c} + 0.81 \ eV; \quad MAE = 0.2 \ eV \tag{15b}$$

$$\Delta E_{O_2^{cc}} = 0.48 \Delta E_{O^c} + 3.03 \ eV; \quad MAE = 0.18 \ eV \tag{15c}$$

$$\Delta E_{Cl(O^c)_2} = 0.41 \Delta E_{O^c} + 2.8 \ eV; \quad MAE = 0.31 \ eV \tag{15d}$$

$$\Delta E_{HO^c} = 0.45 \Delta E_{O^c} - 0.56 \ eV; \quad MAE = 0.22 \ eV \tag{15e}$$

The scaling relations have been shown in the Figure 2 in the main paper. At a given $pH$ and $a_{Cl^-}$, using the scaling relations, we can find the free energy of adsorption of the intermediates (Cl, OH, ClO, O$_2$Cl and O$_2$) at the cus site can be found as a function of the descriptor ($\Delta E_{O^c}$) and potential ($U$). Hence by finding the most stable species on the surface (intermediate with the least free energy of adsorption) at a given $\Delta E_{O^c}$ and U value, we can create the generalized Pourbaix diagram as shown in the Figure 3 in the main paper.

In Figure S4a, we show the scaling between $\Delta E_{O_2^{cc}}$ and $\Delta E_{O^c}$ for 5 randomly chosen functions from the ensemble of functionals. Hence in order to understand the effect of uncertainty in the DFT calculations, we find the uncertainty in the scaling relations and propagate that error in the generalized Pourbaix diagram. The Figure S4b and S4c show the distribution of the slopes and intercepts of scaling between $\Delta E_{O_2^{cc}}$ and $\Delta E_{O^c}$ for the ensemble of functionals. Hence for a each member of the ensemble of functionals, we have a set of scaling relations which lead to a generalized Pourbaix diagram. By creating such ensemble of Pourbaix diagrams and comparing the results with the BEEF-vdW best fit functional, we quantify the confidence value in the predicted surface state as described in the section 2 of

main paper.

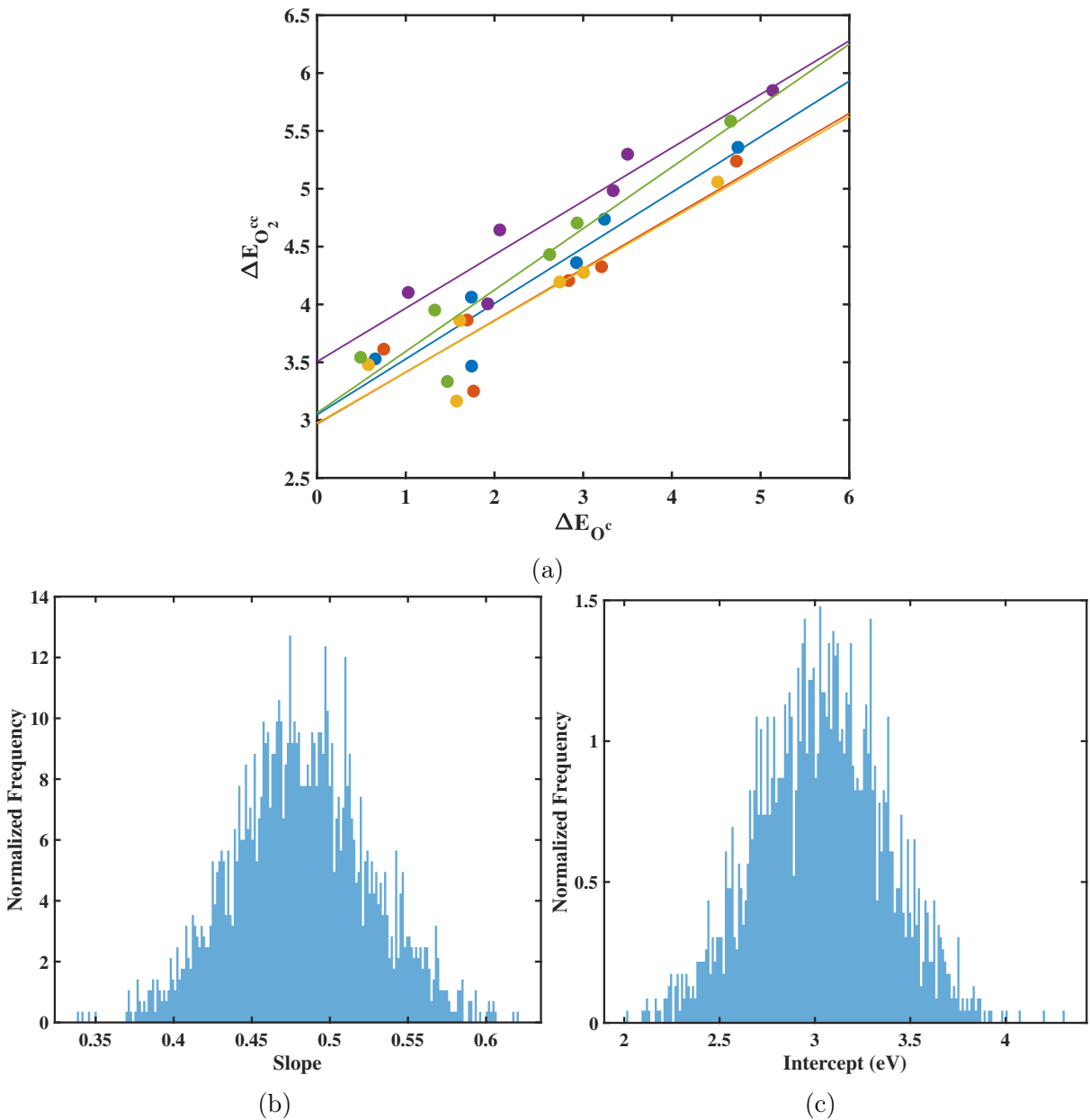

(a)

(b)

(c)

Figure S4: Distributions representing the uncertainty in scaling relations of the interemediate species on rutile oxides of Ru, Ir, Pt, Ti, Sn, V and Rh that are relevant for chlorine evolution. (a) Scaling relations between the adsorption energy of $O_2^{cc}$ vs. the average adsorption energy of O for fully covered surfaces from a set of 5 representative functionals (randomly sampled for demonstration) within the BEEF-vdW XC functional. The adsorption energies and the corresponding scaling relation from each functional are both represented using the same color. The distribution of (b) the slope and (c) the intercept of the scaling lines between $\Delta E_{O_2^{cc}}$ and $\Delta E_{O^c}$ for all the functionals in the ensemble.

Using the scaling relations with $\Delta ZPE$ and $\Delta S$ corrections, the activity volcano for all

10

the three reaction mechanism considered can be expressed as a function of the descriptor, $\Delta E_{O^c}$. Hence the limiting potential for the reaction mechanisms (for the BEEF-vdW best-fit functional) can be expressed as:

I. **Pathway mediated by the intermediate ClO$^c$**

$$U_L = U^{eq} + |\Delta G(ClO^c) - \Delta G(O^c)|/e$$

$$= U^{eq} + |(0.4\ \Delta E_{O^c} + 0.81 + 0.05 + 0.37) - (\Delta E(O^c) + 0.05)|/e$$

$$= U^{eq} + |-0.6\ \Delta E_{O^c} + 1.18|/e \qquad (16)$$

II. **Pathway mediated by the intermediate Cl(O$^c$)$_2$**

$$U_L = U^{eq} + |\Delta G(Cl(O^c)_2) - \Delta G(O_2^{cc})|/e$$

$$= U^{eq} + |(0.41\Delta E_{O^c} + 2.8 + 2 \times 0.05 + 0.37) - (0.48\Delta E(O^c) + 3.03 + 2 \times 0.05)|/e$$

$$= U^{eq} + |-0.07\ \Delta E_{O^c} + 0.14|/e \qquad (17)$$

III. **Pathway mediated by the intermediate Cl$^c$**

$$U_L = U^{eq} + |\Delta G(Cl^c)|/e$$

$$= U^{eq} + |0.34\ \Delta E(O^c) - 1.62 + 0.37|/e$$

$$= U^{eq} + |0.34\ \Delta E(O^c) - 1.25|/e \qquad (18)$$

These activity volcanoes have been shown in the Figure 4 of the main paper. We plot the activity volcano overlapping the Pourbaix diagram to ensure that the reaction mechanism that is active for a given $\Delta E_{O^c}$ is the one governing the limiting potential — giving rise to the generalized activity volcano. Since there exist an ensemble of such scaling relations, we generate an ensemble of generalized activity volcanoes. Using the methodology discussed in the Section 2.3 in the main paper, we find the distribution of $U_L$ and the corresponding probability weighted average of the distribution — expected limiting potential ($U_{EL}$) (Figure 4 in the main paper).

# 6 Quantifying agreement between functionals for specific intermediates

By definition, c-value compares the prediction of the various functionals to that of the BEEF-vdW optimal functional. In order to get a comparison between the various functionals, we define the c-value, $c_{s_i}(U, \Delta E_{O^c})$, of a specific state of the surface, $s_i$, and can be defined as the fraction of the functionals that predict $s_i$ to have the lowest Gibbs free energy. It can be mathematically expressed as follows[2]:

$$c_{s_i}(U, \Delta E_{O^c}) = \frac{1}{N_{ens}} \sum_{n=1}^{N_{ens}} \prod_{j \neq i} \Theta(\Delta G^n_{s_j}(U, \Delta E_{O^c}) - \Delta G^n_{s_i}(U, \Delta E_{O^c})) \qquad (19)$$

where, the summation is over the $N_{ens}$ ensembles, and the product is over all the surface states except the surface state in consideration. The $c_{s_i}$ calculated for the various intermediates ($O^c$, $ClO^c$, $O_2^{cc}$, $Cl(O^c)_2$, $Cl^c$) and Clean surface as a function of the material descriptor $\Delta E_{O^c}$ and U is plotted in Figure S5

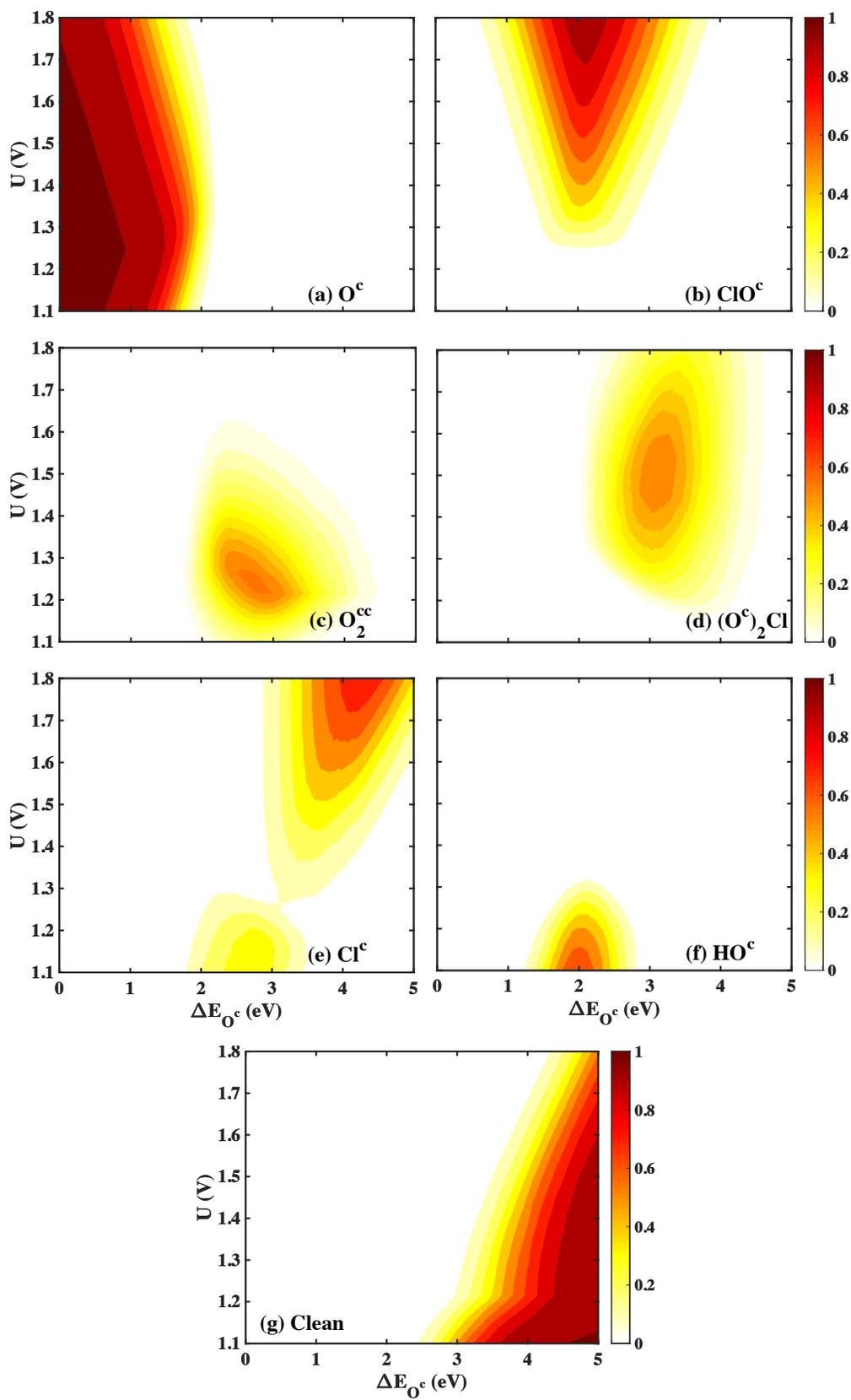

Figure S5: Contour maps showing quantified confidence in the surface predominance diagrams of various intermediates on the catalyst surface based on the generalized Pourbaix diagram as a function of the material descriptor and the potential: (a) $O^c$, (b) $ClO^c$, (c) $O_2^{cc}$, (d) $Cl(O^c)_2$, (e) $Cl^c$, (f) $OH^c$, (g) clean surface.

13

Figure 3 in the main paper, gives an understand of the predictions of the BEEF-vdW optimal prediction and the uncertainty associated with those predictions. Figure S5 helps understanding the uncertainty in predicting a given intermediate on the surface using the uncertainty in the scaling relations. Hence, for a given material (specific $\Delta E_{O^c}$ value), using the Figure S5, one can find the state of the surface as a function of U, and determine how many functionals agree with the predictions. In Figure S6, we show this for $\Delta E_{O^c} = 3.0$ eV.

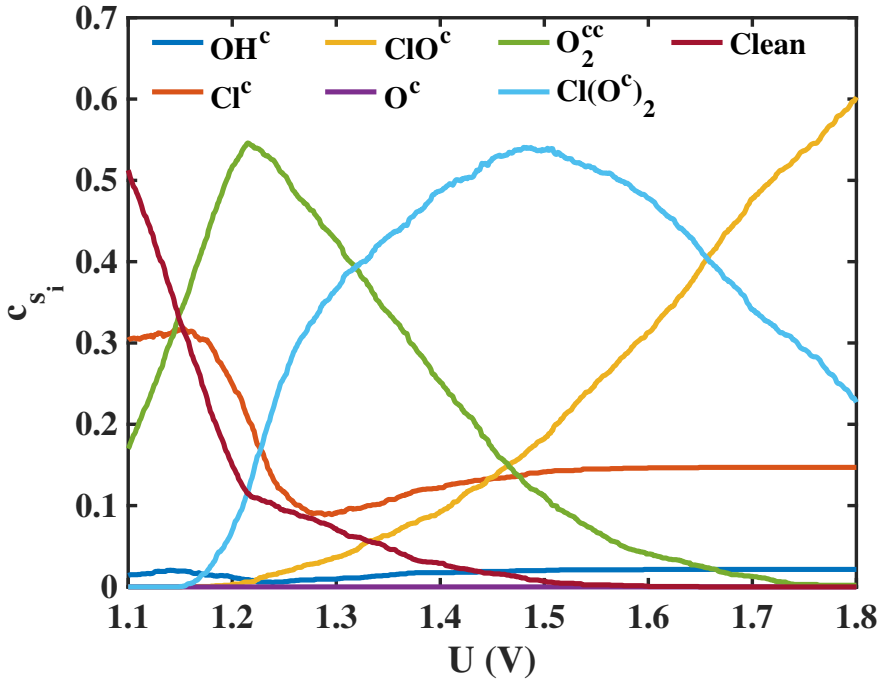

Figure S6: Quantifying confidence in the predicted surface states of the surface, $c_{s_i}$, shown for at $\Delta E_O^c = 3.0$ eV (Refer section 2.4 of the main paper).

From the Figure S6, for a material with $\Delta E_{O^c} = 3$ (eV), at pH=0, we see that majority of functionals predict that surface is clean for $\sim$U<1.2 V and with increase in potential, the surface is covered with $O_2^{cc}$ for $\sim$1.2<U<1.35. On further increasing the potential, we see that surface is covered with $Cl(O^c)_2$ and for $\sim$ U>1.7, surface is covered with $OCl^c$.

14

# 7 Confidence in predicted mechanism vs. predicted activity

From Figure 5 in the main paper, we determine that the confidence in the predicted reaction mechanism ($c_{m_{pred}}$) for $RuO_2$ is 1, mediated by $ClO^c$. However, there is a large distribution in the predicted limiting potential as shown in Figure S7.

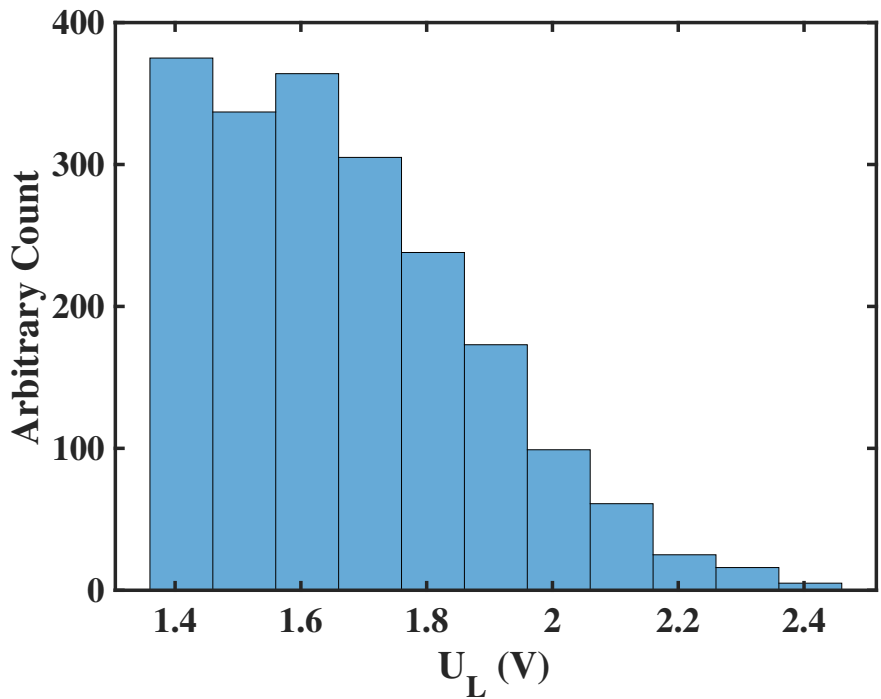

Figure S7: Distribution of the predicted limiting potential ($U_L$) for $RuO_2$ ($\Delta E_{O^c} = 1.53$ eV).



\end{document}